\documentclass[reprint,
superscriptaddress,
 amsmath,amssymb,
 aps,
prb,
]{revtex4-2}

\usepackage{subfig}
\usepackage[utf8x]{inputenc}
\usepackage{graphicx}
\usepackage{dcolumn}
\usepackage{amssymb,amsmath}
\usepackage{bm}
\usepackage{xcolor}
\usepackage{tabularx}
\usepackage{xcolor}
\begin{document}
\title{Magnetization reversal in FePt thin films: experiments and simulations}
\author{A. Román}
\email{augusto.jre@gmail.com}
\author{A. Lopez Pedroso}
\affiliation{Instituto de Nanociencia y Nanotecnología  CNEA-CONICET - Nodo Constituyentes, Av. Gral. Paz 1499, 1650, San Martín, Pcia. de Buenos Aires, Argentina}

\author{K. Bouzehouane}
\affiliation{Unité Mixte de Physique CNRS, Thales, Universite Paris-Saclay, 91767 Palaiseau, France}

\author{J. E. Gómez}
\author{A. Butera}
\affiliation{Instituto de Nanociencia y Nanotecnología, CNEA - CONICET, Nodo Bariloche, Av. Bustillo 9500, 8400, San Carlos de Bariloche, Río Negro, Argentina}
\affiliation{Laboratorio de Resonancias Magnéticas, Gerencia de Física (GF), Centro Atómico Bariloche, CNEA \& Instituto Balseiro Universidad Nacional de Cuyo, Av. Bustillo 9500, 8400, San Carlos de Bariloche, Río Negro, Argentina}

\author{M. H. Aguirre}
\affiliation{Instituto de Nanociencia y Materiales de Aragón, INMA-CSIC-Universidad de Zaragoza, E-50018 Zaragoza, Spain}
\affiliation{Departamento de Física de la Materia Condensada, Universidad de Zaragoza, E-50009 Zaragoza, Spain}
\affiliation{Laboratorio de Microscopías Avanzadas, Universidad de Zaragoza, E-50018 Zaragoza, Spain}

\author{M. Medeiros Soares}
\affiliation{Laboratório Nacional de Luz Síncrotron (LNLS), Centro Nacional de Pesquisa em Energia e Materiais (CNPEM), 13083-970 Campinas, São Paulo, Brazil}
\affiliation{Departamento de Física, Universidade Federal da Paraíba, 58051-900 João Pessoa, Brazil}

\author{C. Garcia}
\affiliation{Departamento de Física and Centro Científico Tecnológico de Valparaíso-CCTVal, Universidad Técnica Federico Santa María, Av. España 1680,  Valparaíso, Chile}

\author{L. B. Steren}
\affiliation{Instituto de Nanociencia y Nanotecnología  CNEA-CONICET - Nodo Constituyentes, Av. Gral. Paz 1499, 1650, San Martín, Pcia. de Buenos Aires, Argentina}
\date{\today}
\begin{abstract}

The competition between shape and perpendicular magnetic anisotropies (PMA) in magnetic thin films gives rise to unusual magnetic behaviors. In ferromagnetic films with moderate PMA the magnetic domain configuration transitions from planar to stripe-like domains above a critical thickness, $t_c$. In this article, we present a detailed study of the magnetization switching mechanism in FePt thin films, where this phenomenon is observed. Using micromagnetic simulations and experiments, we found that below $t_c$ the reversal mechanism is well described by the two-phase model while above this thickness the magnetization within each stripe reverses by coherent rotation. We also analyzed the PMA and its temperature dependence, probing that substrate-induced strains are responsible for the abnormal coercive field behavior observed for FePt films with $t>t_c$.

\end{abstract}
\maketitle

\section{\label{sec:introduction}Introduction}

The magnetization reversal mechanisms in thin films
have been intensively studied in recent years due to its implications for understanding hysteresis loops and technological applications \cite{mathews2010magnetization}. The use of magnetic components in random access memories, for example, requires that they show fast and replicable magnetization switching \cite{yoo2003switching}. Therefore, the analysis of the reversal magnetization mechanisms and the possibility of controlling them are an essential input for the evaluation and design of magnetic materials for information storage \cite{wang2016magnetization, hauet2014reversal, yoo2003switching}. 

Magnetic thin films with striped-domain configuration have potential applications in devices \cite{spain1966stripe} and sensors \cite{garnier2020stripe}, among others. Moreover, recent investigations show the possibility of controlling the transmission of spin waves using striped domains \cite{sadovnikov2017toward,liu2019current}.

The domain structure of these materials is defined by the competition between a perpendicular magnetic anisotropy $K_{PMA}$ and the shape anisotropy $2\pi M_S^2$ where $M_S$ is the saturation magnetization. For a quality factor $Q=\frac{K_{PMA}}{2\pi M_S^2} $ smaller than 1,  the domain structure of these films depends critically on their thickness: above some critical thickness, $t_c$, the out-of-plane magnetization component is periodic and organized in stripes with a sine-like profile, while below $t_c$ the
magnetization lies in the plane of the films. The striped-domain configuration of these systems has been called $weak$ $stripes$ to differentiate it from the striped-domain configuration that appears for $Q>1$, where stripes occur even for extremely low thickness with sharp transitions between oppositely magnetized stripes \cite{garnier2020stripe}. 

The weak stripe domain configuration have been observed in a wide variety of thin films, e.g.  Ni$_{80}$Fe$_{20}$\cite{camara2017magnetization}, Fe$_{1-x}$Ga$_x$\cite{fin2015plane},  Fe-N\cite{garnier2020stripe} and FePt\cite{alvarez2015tunable,guzman2013abnormal, leva2010magnetic} thin films. In these compounds, the PMA and, consequently, the critical thickness has been tuned by changing substrates or adjusting the alloy concentration \cite{tacchi2014rotatable}.

Epitaxial FePt thin films have been extensively investigated for perpendicular magnetic recording media applications due to their high perpendicular magnetic anisotropy.  \cite{spada2003x,toney2003thickness}. The crystal structure of bulk and epitaxial FePt thin films is a chemically ordered body-centered tetragonal cell (L1$_0$) \cite{lyubina2007nanocrystalline,bayliss1990revised,yuasa1994magneto, soares2011highly,soares2014orbital} at room temperature. This structure can also be described using a face-centered tetragonal (fct) pseudo-cell for which the following lattice parameters have been reported: $a_{\mathrm{FePt}}=3.852$ \AA \ and  $c_{\mathrm{FePt}}=3.713$ \AA. The tetragonality of this pseudo-cell is $\frac{c}{a}=0.964$ \cite{lyubina2007nanocrystalline,hai2003original, buschow1983magneto}.
   
In thin films, the formation of the L1$_0$  phase requires high temperatures (\textgreater 400 $^o$C) during the fabrication process or post-deposition treatments, which lead to large grains unsuitable for magnetic recording \cite{toney2003thickness,white2000physical}. Films deposited at room temperature generally form disordered alloys crystallizing in the A1 fcc crystalline structure with small grain sizes \cite{leva2010magnetic}. The reported PMA in these films is $\sim 1\  \frac{\mathrm{Merg}}{\mathrm{cm}^3}$, almost two orders of magnitude smaller than that of the  ordered FePt films \cite{guzman2013abnormal}. A1 FePt thin films present a transition between in-plane domains to stripe-like domains that occurs at a critical thickness $t_c \approx 30$ nm \cite{guzman2013abnormal}. Guzmán and co-workers \cite{guzman2013abnormal} analyzed the temperature dependence of the magnetization of A1 FePt films and reported an abnormal behavior of the coercivity. The authors associated this behavior with a transition from stripe-type to in-plane domains.

The magnetization reversal mechanism in FePt films is still a matter of controversy. In this article, we present experiments and micromagnetic simulations aiming to deepen the understanding of the magnetic configuration of Q \textless 1 thin films and hence their magnetization reversal processes. The magnetization loop parameters, and their correlation with the system anisotropies and crystalline structure will also be discussed.

\section{Magnetization reversal models}
The analysis of the magnetization reversal will be performed in the frame of three models: (I) Coherent rotation, (II) domain wall motion, and (III) the two-phase model. The first two models represent extreme cases of switching behavior, while the third one results from the combination of the first two cases \cite{coffey2002angular, mathews2010magnetization}. The angular dependence of the coercivity (ADC) has a characteristic behavior in each model. The ADC is thus an excellent prove for analyzing the magnetization reversal mechanisms in magnetic materials \cite{oh2005crystallographic}. 

Stoner and Wohlfarth \cite{stoner1948mechanism} developed the coherent rotation model for a single domain particle with uniaxial anisotropy. S-W provides also a good description of the magnetization switching by rotation in thin films \cite{coffey2002angular}. The ADC in the S-W model is given by:

\begin{equation}
H_C(\varphi)=H_0
\left\{
\begin{array}{cc}
     \frac{1}{\left(\cos(\varphi)^\frac{2}{3}+\sin(\varphi)^\frac{2}{3}\right)^\frac{3}{2}}\ , &  0<\varphi\leq\frac{\pi}{4}\\
     \frac{\sin(2\varphi)}{2}\ , & \frac{\pi}{4}\leq\varphi\leq\frac{\pi}{2}
\end{array}
\right.,
\label{Hc_SW}    
\end{equation}

where $\varphi$ is the angle between the direction of the applied field and the easy axis and $H_0$ is the coercive field when the magnetic field is applied along the easy axis.

The Kondorsky formula (Equation \ref{kondorskyformula}) predicts the ADC of the magnetization reversal mechanism by domain wall movement \cite{kondorsky1940hysteresis}. In this case, the magnetization reverses when the Zeeman contribution overcomes the domain wall energy. This mechanism has been observed in many magnetic thin films \cite{byun1986study, shtrikman1959coercive, fisher1990switching, jeong2000magnetic}:

\begin{equation}
  H_C\left(\varphi\right) = \frac{H_0}{\cos\left(\varphi\right)}.
  \label{kondorskyformula}
\end{equation}

Suponev \cite{suponev1996angular} \textit{et al} generalized the Kondorsky model, proposing a two-phase model. This model assumes that there are only two types of magnetic domains, e.g., two phases. The magnetization of the whole system reverses either by coherent rotation or domain wall movement depending on the magnetic field range. In reference \citenum{suponev1996angular}, the ADC for an ellipsoid of revolution with a uniaxial anisotropy along the $y$ axis is thus deduced: 

\begin{equation}
    H_C\left(\varphi\right) = \frac{H_0\cos\left(\varphi \right)}{\frac{1}{y}\sin^{2}\left(\varphi\right)+\cos^{2}\left(\varphi\right)},  y =\frac{N_{A}+N_{x}}{N_{y}}.
    \label{twophasemodel}
\end{equation}
$N_x$, $N_y$, and $N_z$ are the demagnetizing factors of the ellipsoid along its main axes, being $N_x=N_z$. $N_A$ is an effective demagnetizing factor that takes into account the contributions of anisotropies other than shape anisotropy favoring an $y$ easy-axis. For an infinite thin film, the demagnetizing factor along the in-plane axes should be $N_x=N_y=0$, which makes $y\rightarrow\infty$ and, as a consequence, reduces the expression to the Kondorsky formula (Equation \ref{kondorskyformula}).

\section{\label{sec:experimental}Experimental}
We studied the magnetism of a series of FePt thin films fabricated by dc magnetron sputtering on naturally oxidized Si (100) substrates. The chamber was pumped down to a base pressure of $1\times10^{−6}$ Torr, and the films were sputtered at 2.6 mTorr of Ar pressure. A power of 20 W, and a target-substrate distance of about 10 cm were used. The sputtering rate was 0.19 $\frac{\mathrm{nm}}{\mathrm{s}}$ for the FePt deposition. A 4 nm-thick Ru layer capped the samples to prevent oxidation.

The film thickness $t$ was varied from 10 nm to 60 nm and checked by X-ray reflectometry. High-resolution synchrotron X-ray diffraction experiments were performed on the XRD2 beamline ($E$=7.00375 keV) at the Laboratorio Nacional de Luz Sincrotron (Campinas, Brazil) using different geometries to collect diffraction patterns. The stacking, interfaces, and crystallinity of the films were analyzed by high-resolution scanning transmission electron microscopy (HRSTEM). High resolution transmission electron microscopy performed by FEI Tital 80-300keV image corrected. 
 
 The magnetic characterization of the samples was made by measuring magnetization loops using a Vibrating Sample Magnetometer (VSM) and by Magnetic Force Microscopy (MFM) to image magnetic domains at the nanoscale. For the VSM measurements, we applied an external magnetic field in the plane of the films, varying the angle between the applied field and the Si[100] direction in the range $0^o<\varphi<180^o$. The magnetization loops were measured between 50 and 300 K. MFM images were recorded by an Asylum AFM, using the phase detection mode at 77 K and 300 K. We used Asylum High Coercive HC5SP3 (H$_c\approx5000$ Oe) tips for the measurements. MFM images of remanence states were measured after saturating the films with a magnetic field of $H\sim6000$ Oe applied in the film's plane.
 
 Micromagnetic simulations using Mumax3 open source software \cite{vansteenkiste2014design, Lel2014, Leliaert2017, Exl2014} were performed to evaluate the magnetic configurations and their dependence with magnetic field. The simulations were made keeping fixed the saturation magnetization,   $M_S=1130$ $\frac{\mathrm{emu}}{\mathrm{cm}^3}$ and  the exchange stiffness constant $A_{ex}=9.5\times10^{-7}$ $\frac{\mathrm{erg}}{\mathrm{cm}}$ \cite{okamoto2002chemical,kanazawa2000magnetic}. To simulate the grains of the film, we defined regions through the Voronoi Tessellation and reduced the exchange interaction between grains, varying its value to reproduce the experimental remanence and coercive field. The perpendicular magnetic anisotropy constant $K_{PMA}$ was varied between 0.7 $\mathrm{\frac{Merg}{cm^3}}$ and 1.8 $\mathrm{\frac{Merg}{cm^3}}$. The calculations were performed at zero temperature. It is expected to get similar results between the simulation and experimental results obtained at room temperature due to the Curie temperature of A1  FePt films is around 580K \cite{weller2016fept}.

\section{Results and discussion}

\subsection{FePt films structural analysis}
\begin{figure}[ht]
\centering
\includegraphics[width=\textwidth]{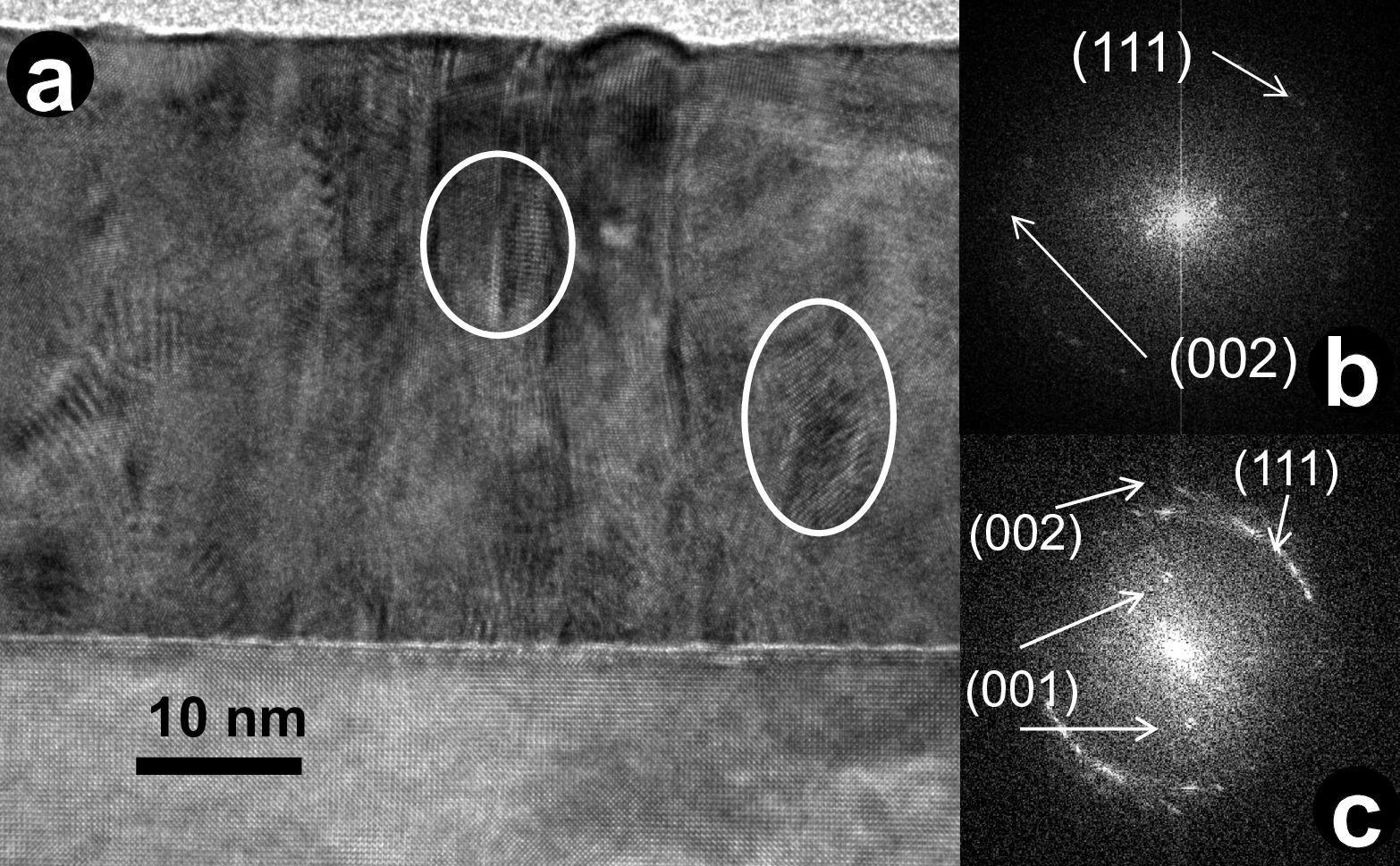}

\caption{\label{FePt_TEM}(a) Cross sectional HR-TEM image and (b) SAED pattern from a 40 nm thickness FePt film deposited on a silicon substrate. White circles indicate ordered regions of the sample}
\end{figure}

We investigated the FePt films crystalline structure by high-resolution transmission electron microscopy (HRSTEM) and X-ray diffraction (XRD). The cross-section of a 40 nm FePt film measured by HRSTEM, shown in Figure \ref{FePt_TEM} (a), puts in evidence the polycrystalline structure of the samples.
The films are formed by grains that present different degrees of crystalline order, e.g. white circles indicate the regions with  highly ordered structure. HRSTEM images also served us to get a description of the film stacking. Films are composed of small grains of a few nanometers (3-4 nm) at the first layers from the substrate,
acquiring a columnar profile with a width of tens of nanometers beyond the interface zone. The average grain size of the sample of 40 nm FePt thickness is $11.3\pm0.4$ nm, where the average was calculated including grains and columns widths.

Figure \ref{FePt_TEM} also shows Fast Fourier Transformed for the calculation of the Selected Area Electron Diffraction patterns of different regions of the sample. Patterns presenting only the (111) reflection were found in some zones of the films and were associated with the A1 disordered phase (Figure \ref{FePt_TEM} (b)). Other regions presented the (001) reflection (Figure \ref{FePt_TEM} (c)), indicating that there are grains formed in the ordered phase (L1$_0$) \cite{okamoto2002chemical}. 

\begin{figure}[ht]
\centering
\subfloat{\includegraphics[width=0.5\textwidth]{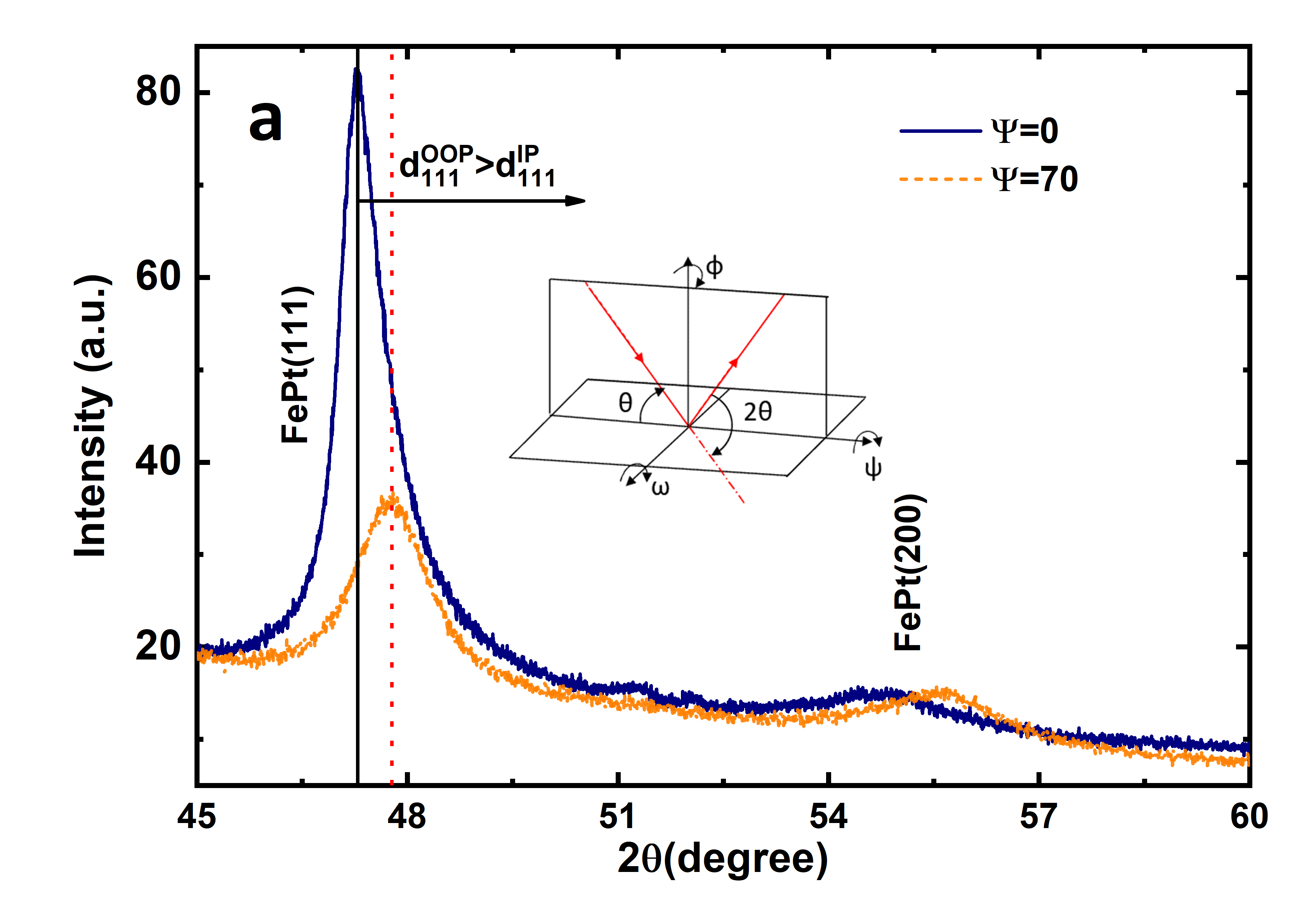}}
\subfloat{\includegraphics[width=0.5\textwidth]{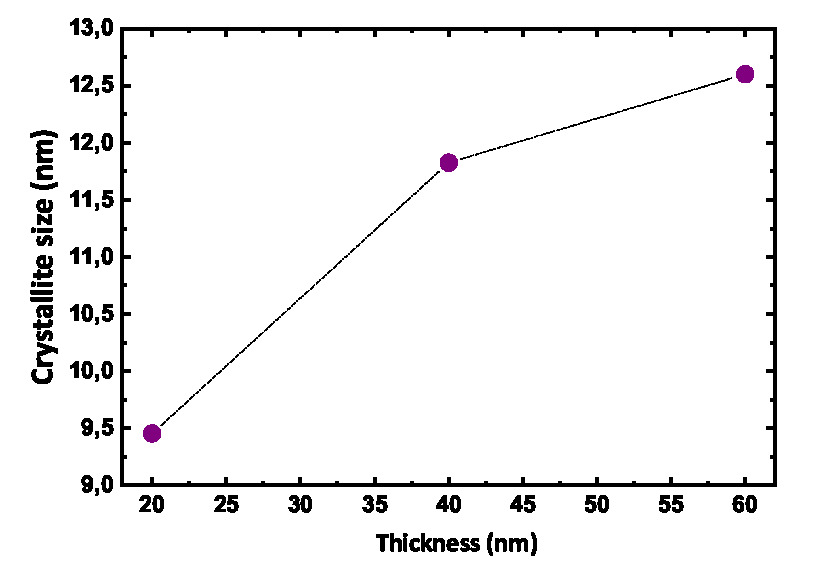}}
\caption {\label{XRD_PSI_RT}(a) XRD patterns of FePt(60nm)/Si recorded for $\theta/ 2\theta$ for $\psi =0$ and $\psi = 70$. (b) Calculated average crystallite size using the Scherrer equation and the width of the (111) reflection vs the film thickness for FePt on silicon.}
\end{figure}

In Figure \ref{XRD_PSI_RT} (a) we show the XRD pattern for a 60 nm-thick FePt film measured using different geometries. On the one hand, we used the conventional $\theta/2\theta$ geometry to measure the distance between crystallographic planes parallel to the substrate surface. On the other hand, we set $\psi = 70^o$ (Figure \ref{XRD_PSI_RT} (a)) in order to measure the distance between planes that are almost perpendicular to the substrate surface \cite{birkholz2006thin_6}. The conventional $\theta/2\theta$ measurement ($\psi=0^o$) indicates that the film presents a (111) texture perpendicular to the film plane. The lattice parameter deduced from the (111) peak is $a=3.85$ \AA\  in agreement with previous works\cite{guzman2013abnormal, leva2010magnetic,ramos2009stripe}. The XRD pattern measured at $\psi=70^o$ presents a shift of the (111) peak to larger angles (Figure \ref{XRD_PSI_RT} (a)), indicating that the substrate induces an in-plane compression on the film. Equation \ref{strain} is an estimation of the film strain along the direction perpendicular to the substrate plane:

\begin{equation}
\epsilon=\frac{2\nu}{(1+\nu)\sin(70^0)}\frac{d_{\psi=0}-d_{\psi=70}}{d_{\psi=0}}\times 100\%.
\label{strain}
\end{equation}

$d_{\psi=0}$  and $d_{\psi=70}$ are the distances between planes parallel to the films surface and those almost perpendicular to the substrate, respectively and $\nu$ is the Poisson's ratio \cite{birkholz2006thin_6, alvarez2015tunable, hsiao2009effect}. Using this expression, we obtained a 0.66\% strain in the FePt film.

From the (111) XRD peak width, we calculated the average crystallite size of the alloy as a function of the film thickness using the Scherrer formula \cite{birkholz2006thin_3} (Figure \ref{XRD_PSI_RT} (b)). 
The crystallite size increases with the film thickness from 9.5 nm to 12.5 nm as expected from the description of the stacking obtained from HRSTEM measurements. Also in the case of the 40 nm FePt sample, the crystallite size measured by XRD (11.8 nm) is in agreement with the value obtained from HRSTEM images.

Takahashi and coworkers \cite{takahashi2004microstructure, takahashi2004size, takahashi2003size} found a strong correlation between the crystalline structure and the grain size of FePt alloys. These authors reported a transition from A1 to L1$_0$ FePt crystalline structure as the grain size varies from 4 nm to 7 nm providing another evidence of the presence of grains in the ordered phase in our samples

The structural characterization reveals two main facts. Firstly, HRSTEM and XRD results suggest a mix of ordered and disordered FePt phases, where the A1 is the major phase with a small fraction of L1$_0$ nanograins embedded; secondly, that the structure of the films is compressed in the plane by the substrate.

\subsection{Critical thickness $t_c$\label{tc_section}}

\begin{figure}[ht]
\centering
\subfloat{\hspace*{-0.4cm}\includegraphics[width=0.5\textwidth]{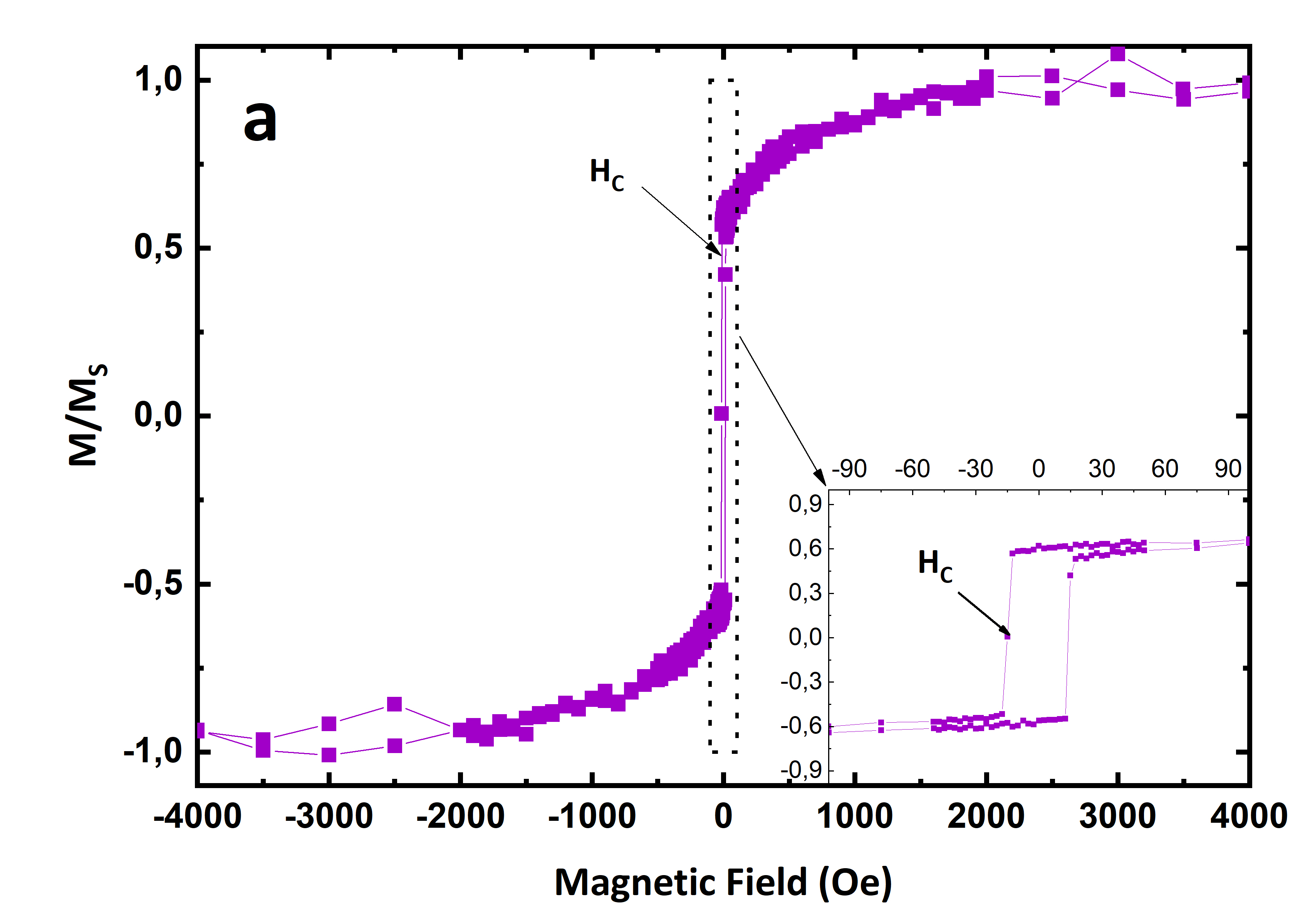}}
\subfloat{\hspace*{-0.4cm}\includegraphics[width=0.5\textwidth]{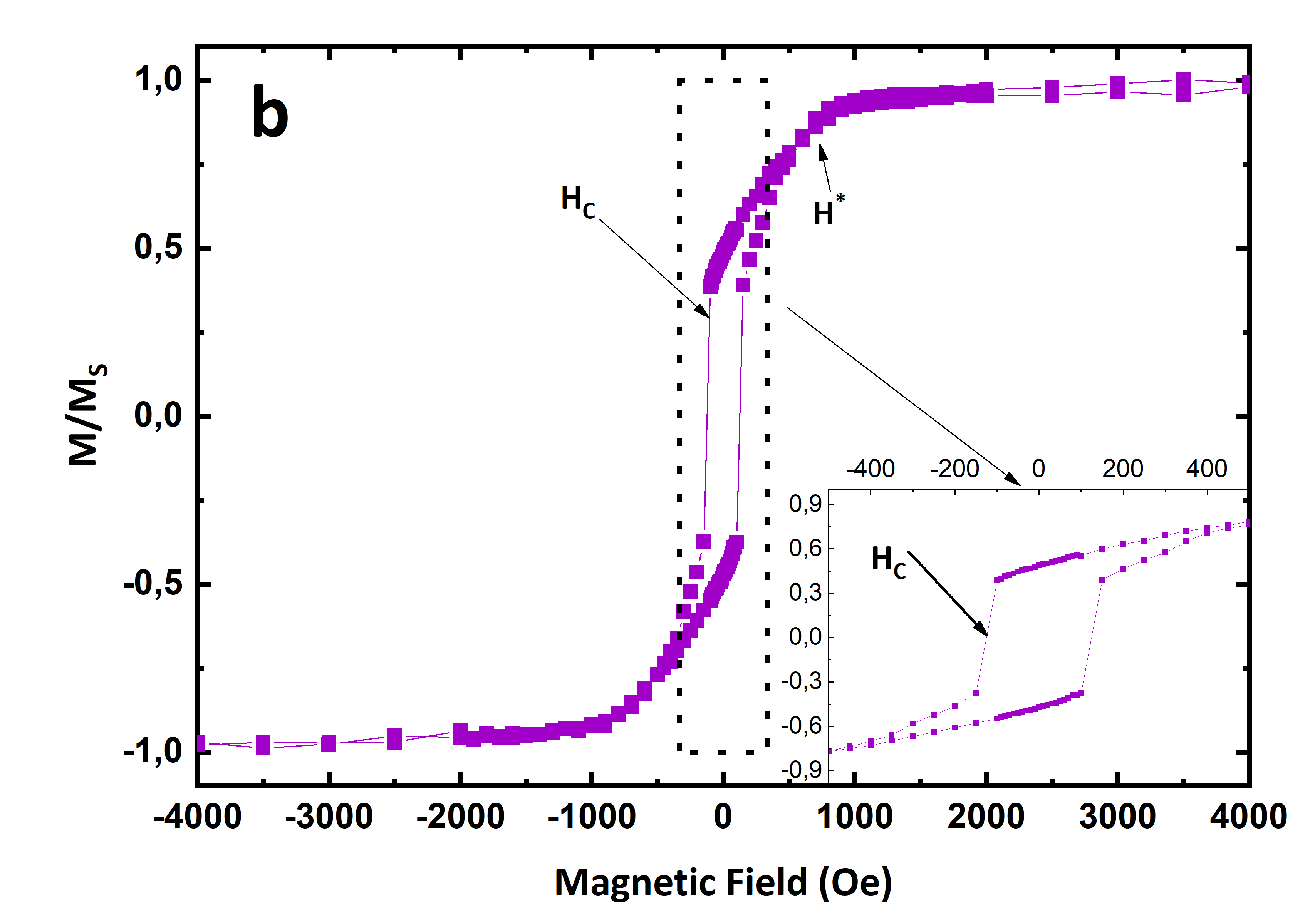}}\\
\subfloat{\includegraphics[width=0.5\textwidth]{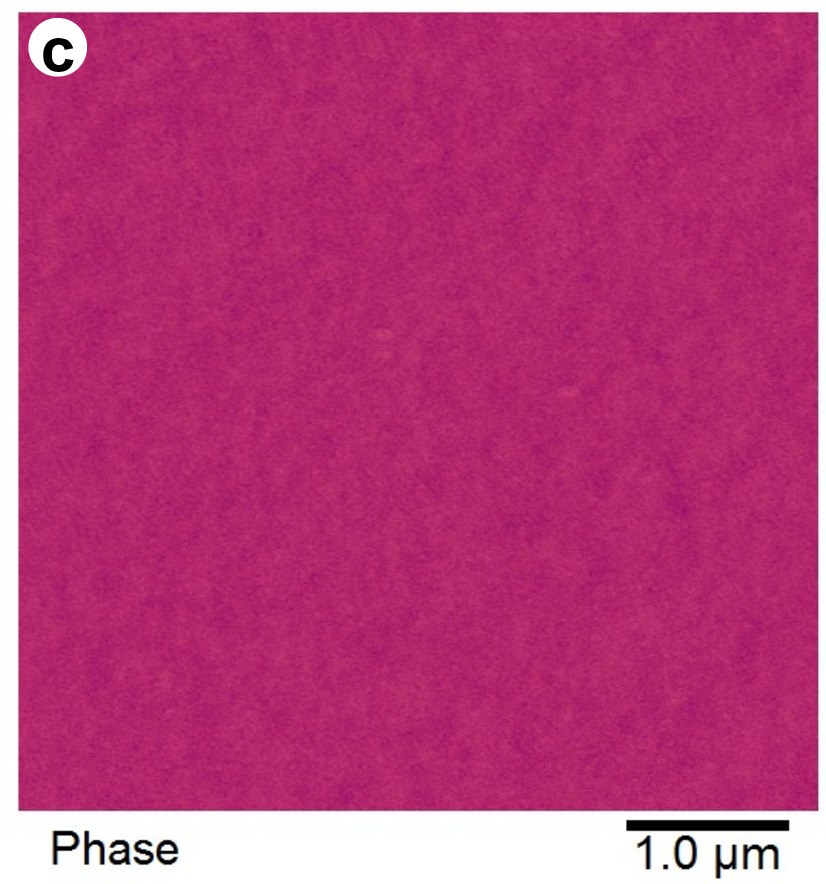}}
\subfloat{\includegraphics[width=0.5\textwidth]{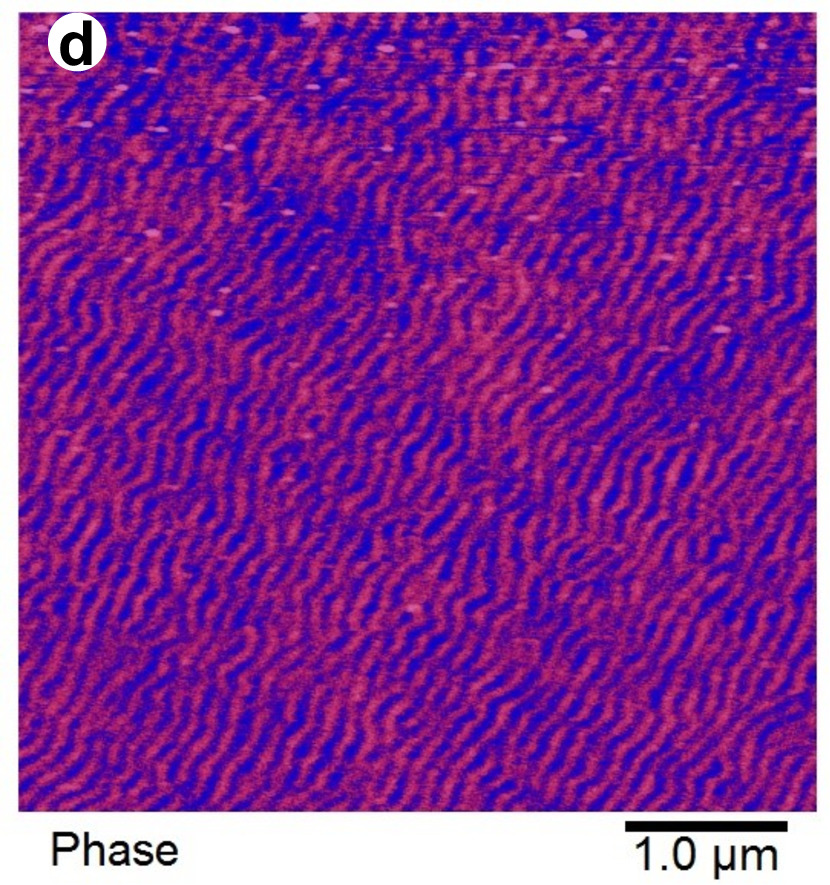}}
\caption{\label{M_H_MFM}In-plane easy axis $(\varphi=0^0)$. hysteresis loops and MFM images for FePt films of (a, c) $t=10$ nm and (b, d) $t=60$ nm, respectively. All the measurements were performed at $T=300$ K. 
}
\end{figure}

The analysis of the hysteresis loops shows a notable difference in the shape of the curves, depending on the film's thickness (Figure \ref{M_H_MFM} (a) and (b)). The magnetization curves for thin FePt films ($t< 30$ nm) present coercive fields smaller than 20 Oe and relatively squared loops (Table \ref{TablaMagnetismo}), while thicker films present larger coercive fields and a linear dependence of the magnetization for $-H_c<H<H^*$ (Figure \ref{M_H_MFM}).

\begin{table}[ht]

\caption{Remanence ($\frac{M_R}{M_S}$), coercive field ($H_C$) and stripes domain period ($\lambda$) of FePt films of different thickness. Uncertainties are shown in parentheses.}

\begin{tabularx}{\columnwidth}{@{\extracolsep{\fill}}c c c c}
\hline
\hline
\begin{tabular}[c]{@{}c@{}}Thickness \\ (nm)\end{tabular} & $\frac{M_R}{M_S}$ & \begin{tabular}[c]{@{}c@{}}$H_c$ \\ (Oe)\end{tabular} &\begin{tabular}[c]{@{}c@{}}$\lambda$ \\ (nm)\end{tabular} \\
\hline
10 & $0.65 (0.02)$  & $15(3)$ &-\\ 
20& $0.84 (0.02)$  & $10(3)$ &-\\
40& $0.62 (0.02)$  & $42(3)$ &$90(10)$ \\ 
49
& $0.53 (0.02)$  & $157(3)$ &$104(10)$ \\
60& $0.46 (0.02)$  & $148(5)$&$110(10)$\\ \hline
\hline

\end{tabularx}

\label{TablaMagnetismo}
\end{table}

In Figure \ref{M_H_MFM} (c) and (d), we present the MFM images of the 10 and 60 nm samples, taken in remanence after saturating the magnetization of the film. The MFM measurements of thin films did not show magnetic contrast; this suggests that the size of the magnetic domains is much bigger than the scan area or that the magnetization is in the plane of the sample. On the other hand, thick films ($t>30$ nm) show striped magnetic domains with out-of-plane magnetization components (Figure\ref{M_H_MFM}(d)). These results indicate a change of the domain structure at a critical thickness $t_c$ between 20 and 40 nm. Our measurements agree with previous results  \cite{leva2010magnetic} that estimated a critical thickness of 30 nm for FePt films deposited on silicon.

Murayama \cite{murayama1966micromagnetics} derived an expression for the stripe period $\lambda$ as a function of  film thickness $t$ and magnetic parameters:

\begin{equation}
    \frac{\lambda}{2}=\sqrt{2\pi t}\sqrt[4]{
\frac{A_{ex}}{2\pi M_S^2}\left(1+\frac{2\pi M_S^2}{K_{PMA}}\right)},
\label{MurayamaLambda}
\end{equation}
where $A_{ex}$ is the exchange stiffness constant, $M_S$ is the saturation magnetization, $K_{PMA}$ is the perpendicular magnetization anisotropy constant and $t$ is the thickness of the film. Replacing the exchange stiffness constant  $A_{ex}=9.5\times10^{-7}\ \mathrm{\frac{erg}{cm}}$ and $M_S=1130\ \mathrm{\frac{emu}{cm^3}}$ from reference \citenum{okamoto2002chemical} in Equation \ref{MurayamaLambda}, we estimated a perpendicular anisotropy constant of $1.4\pm0.5\ \mathrm{\frac{Merg}{cm^3}}$ for films with thickness above $t_c$. This result agrees with the value obtained from the magnetometry measurements using the area-method \cite{johnson1996magnetic} and with the values reported in previous studies on disordered FePt films deposited on silicon\cite{leva2010magnetic,guzman2013abnormal}.

\subsection{Magnetization reversal mechanism at room temperature}

Both the magnetization and the MFM results suggest that the magnetization reversal mechanism should depend on the FePt film thickness, being particularly different for films with $t<t_c$ and  $t>t_c$.  To analyze the samples' magnetization reversal process, we performed measurements of the angular dependence of the magnetization curves with the magnetic field applied in the plane of the films.

\subsubsection{Films with $t<t_c$}

\begin{figure}[ht]
\centering
\subfloat{\includegraphics[width=0.593\textwidth]{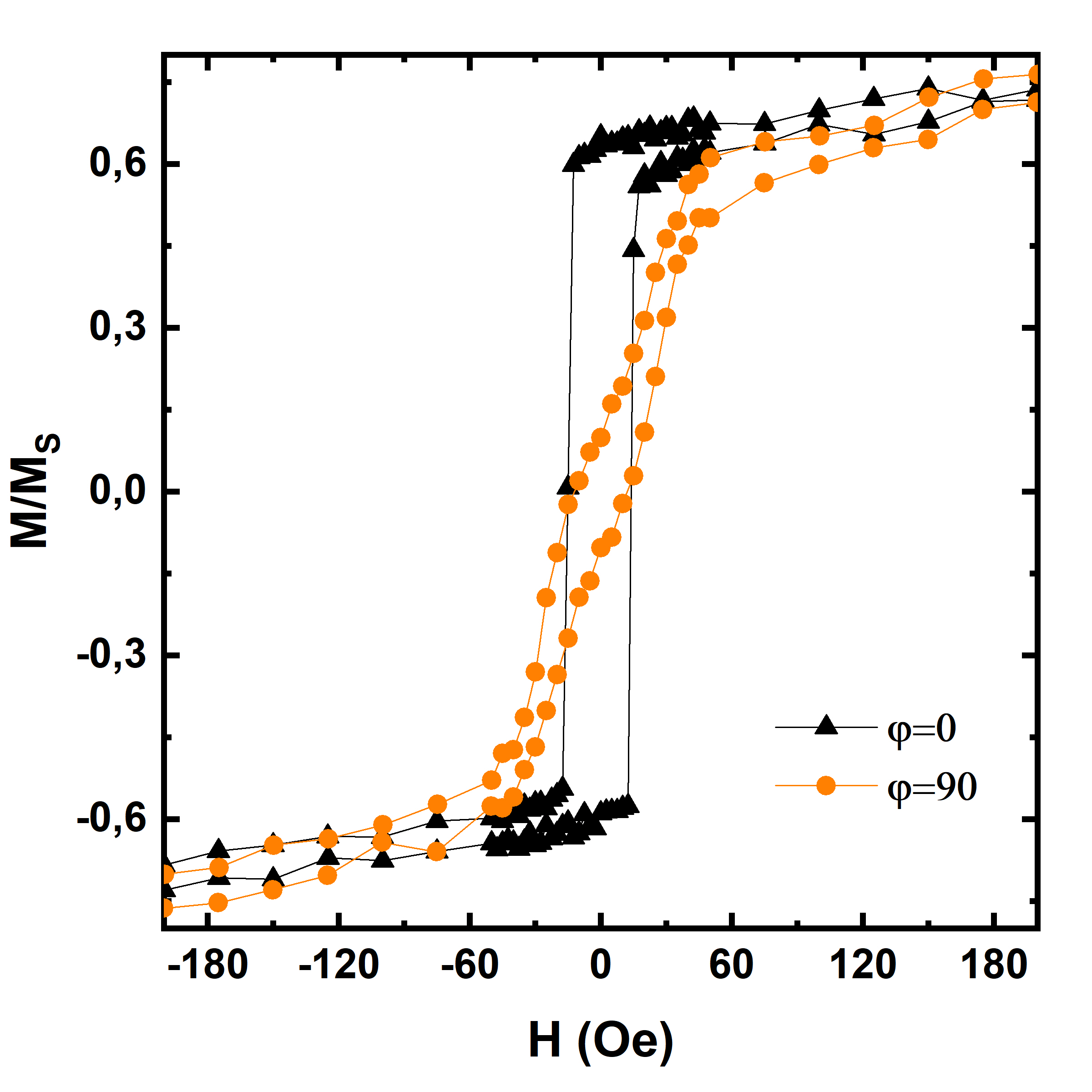}}
\begin{minipage}[b]{0.407\textwidth}
\includegraphics[width=\textwidth]{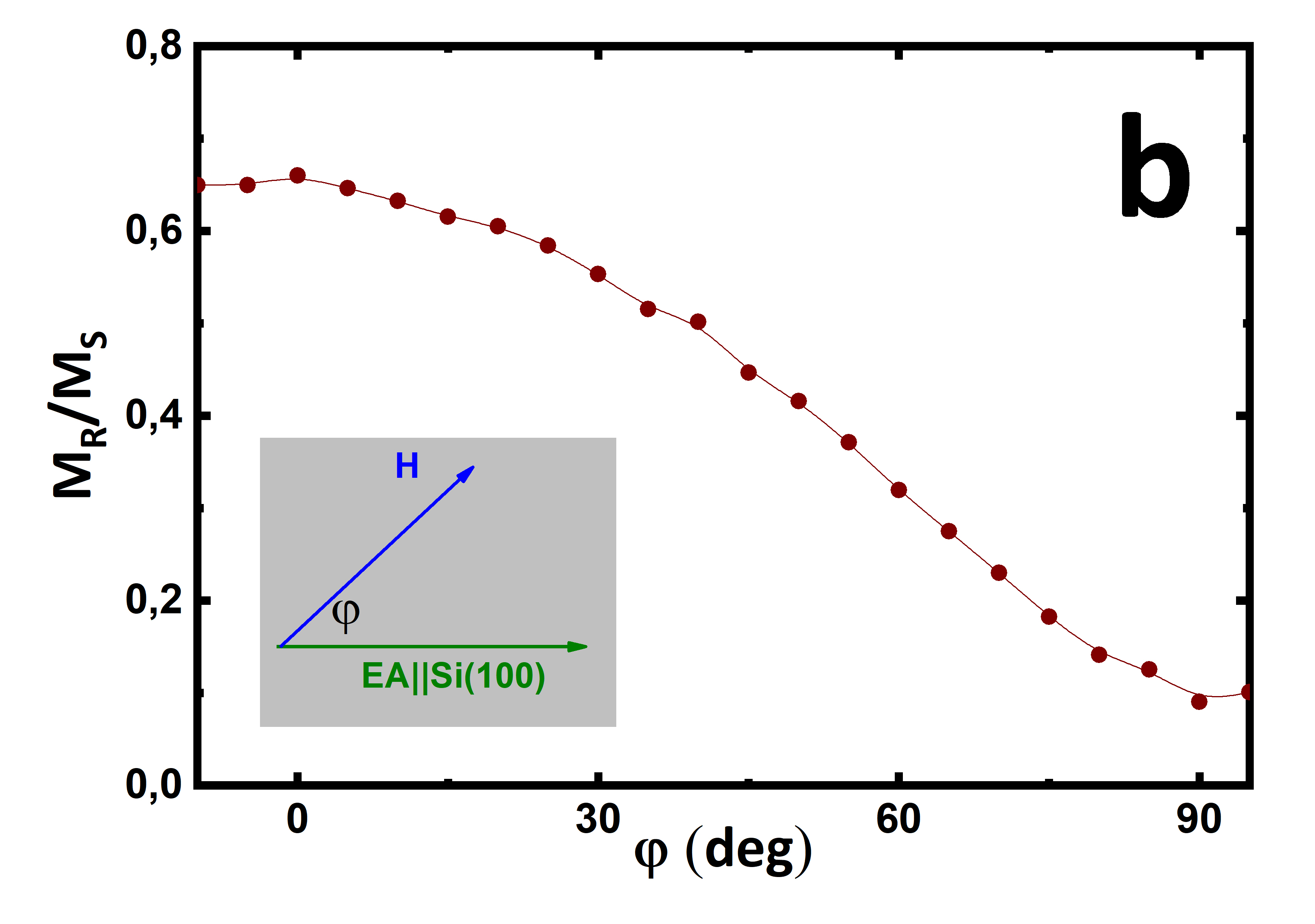}\
\includegraphics[width=\textwidth]{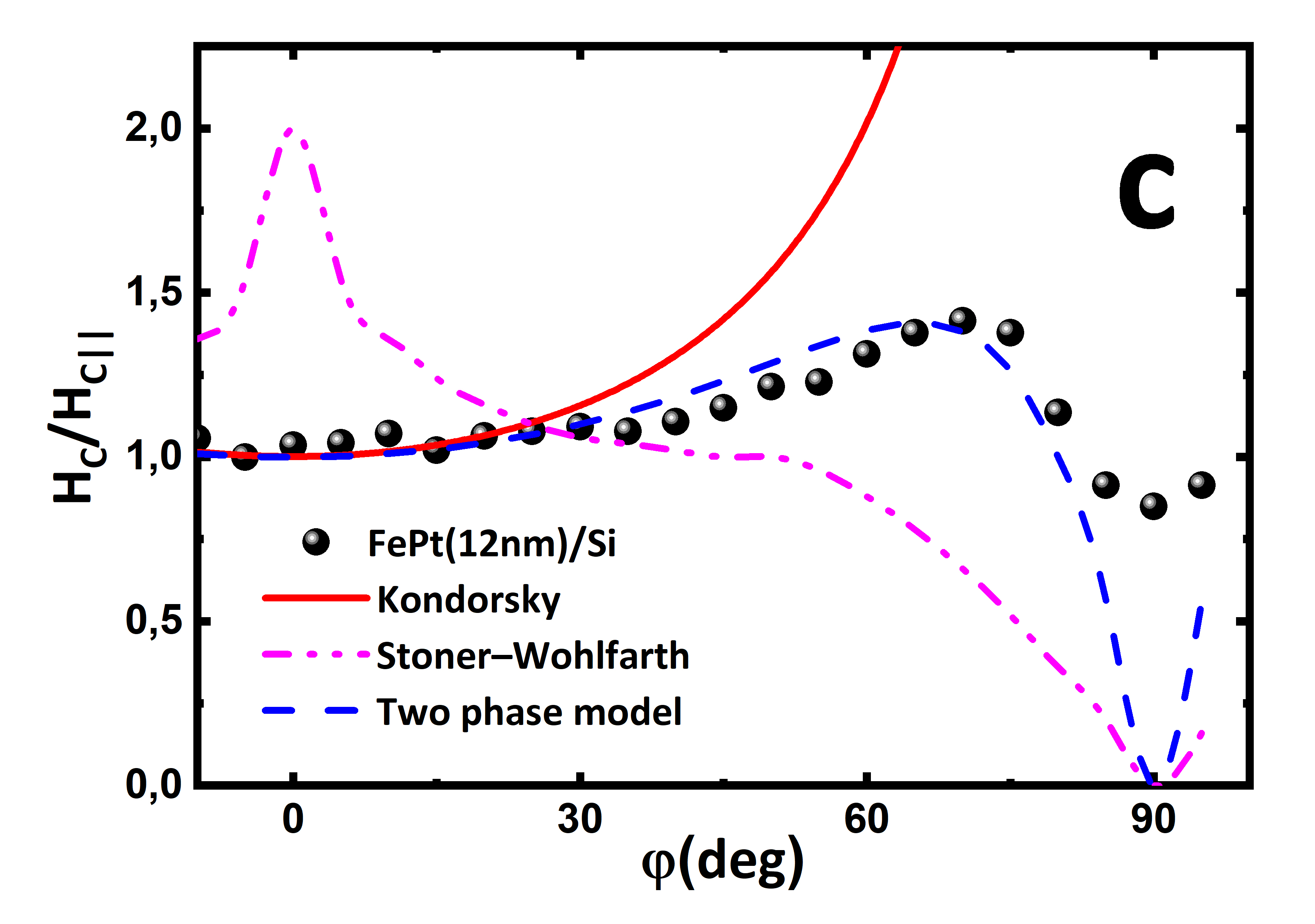}
\end{minipage}
\caption{\label{Mr_angular_10nm}(a) Hysteresis loops measured for $\varphi=0^0$ and $\varphi=90^0$ for FePt(10 nm)/Si. (b)  angular dependence of the remanent magnetization. (c) Angular dependence of the coercive field for a 10 nm thickness FePt thin film plotted together with the magnetization reversal models: Stoner-Wohlfarth, Kondorsky and two phase model.
}
\end{figure}

 The angular dependence of the remanent magnetization for films thinner than $t_c$ reveals the presence of an in-plane uniaxial anisotropy with an easy-axis oriented along the Si[100] direction (Figure \ref{Mr_angular_10nm} (a) and (b)). The magnetization loops measured with the magnetic field applied parallel, ($\varphi=0^o$) and perpendicular ($\varphi=90^o$)  to the easy-axis (e.a) are shown in Figure \ref{Mr_angular_10nm} (a).  A square loop with $H_c=$15 Oe is observed for loops measured $\varphi=0^o$, while a two-steps loop is observed for $\varphi=90^o$. The two-steps loop can be related to films with two anisotropies: a uniaxial induced during the fabrication process or the morphology of the substrate and a biaxial induced by the substrate structure \cite{bisio2006tuning}.

The angular dependence of the coercive field is shown in Figure \ref{Mr_angular_10nm} (c). As can be seen, the coercive field increases with $\varphi$, reaching a maximum value close to  $\varphi=90^o$. In the same figure, the experimental data was compared with calculated values arisen from Stoner-Wohlfarth\cite{stoner1948mechanism}, Kondorsky \cite{kondorsky1940hysteresis}, and two-phase models \cite{suponev1996angular}. The two-phase model \cite{suponev1996angular} is the best fit for our experimental data,  besides the fact that the coercive field is not zero at the hard axis, as expected for this model. Suponev and coworkers \cite{suponev1996angular} attributed the non-zero Hc($\varphi=0^o$) to the grain easy axis distribution due to the polycrystalline nature of the sample. 

The presence of a relatively strong perpendicular magnetic anisotropy \cite{guzman2013abnormal} induced by strain was reported for polycrystalline FePt thin films and could explain the finite value of $y= 6.76$ derived for our films using  Equation \ref{twophasemodel}. 

\subsubsection{Films with $t>t_c$\label{Reversal_above_tc}}

\begin{figure}[ht]
\centering
\subfloat{\includegraphics[width=0.45\textwidth]{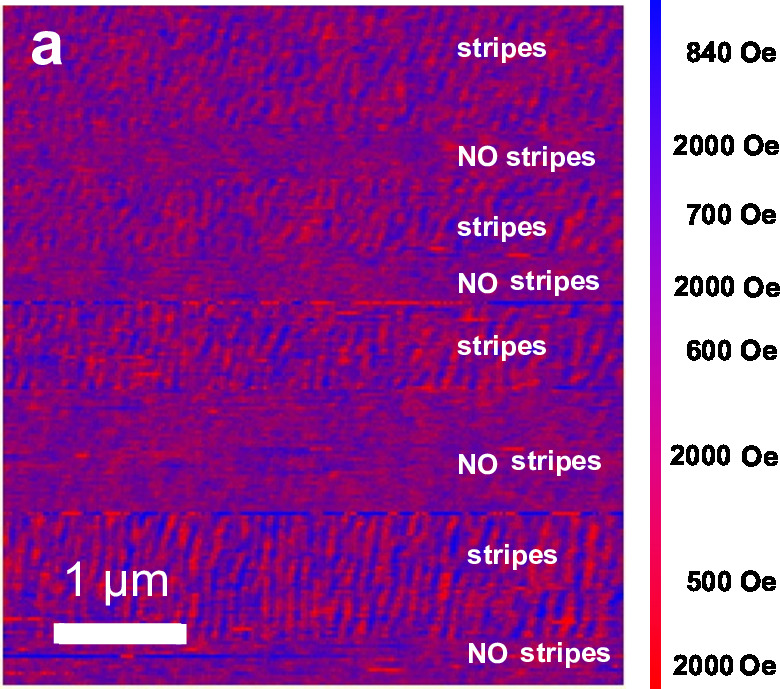}}
\subfloat{\includegraphics[width=0.54\textwidth]{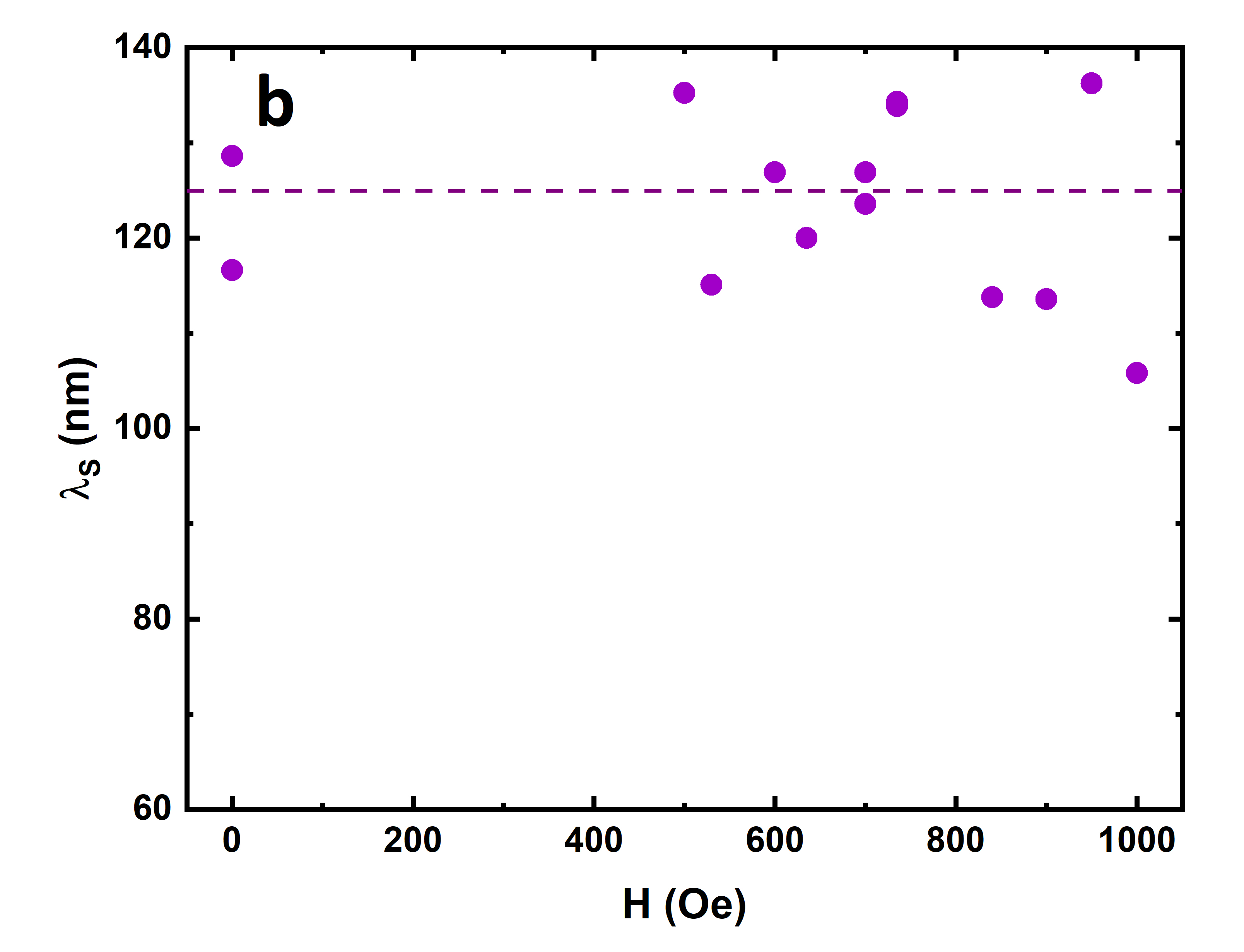}}
\caption{\label{MFM_vs_H} (a) MFM image of a representative FePt film with an external magnetic field applied along the direction parallel to the stripes. (b) Stripe-domains period ($\lambda_S$) calculated from MFM images as a function of the magnetic field.
}
\end{figure}

The magnetization loops for thick films do not depend appreciably on the magnetic field orientation which is expected due to the rotatable anisotropy that characterizes films with stripe-domains\cite{alvarez1997perpendicular,leva2010magnetic, garnier2020stripe}. 
This behavior arises from the stripe-domain structure of these films at remanence. Therefore, complementary experiments and micromagnetic simulations were needed to get an insight into the magnetization reversal mechanisms of these films. MFM measurements were performed under an external magnetic field applied along the direction of the stripes. The stripe patterns were observed up to 1000 Oe and disappeared for magnetic fields larger than 2000 Oe (Figure\ref{MFM_vs_H} (a)). However, the stripe period does not depend appreciably on the magnetic field intensity (Figure\ref{MFM_vs_H} (b)).

\begin{figure}[ht]
\centering
\includegraphics[width=\textwidth]{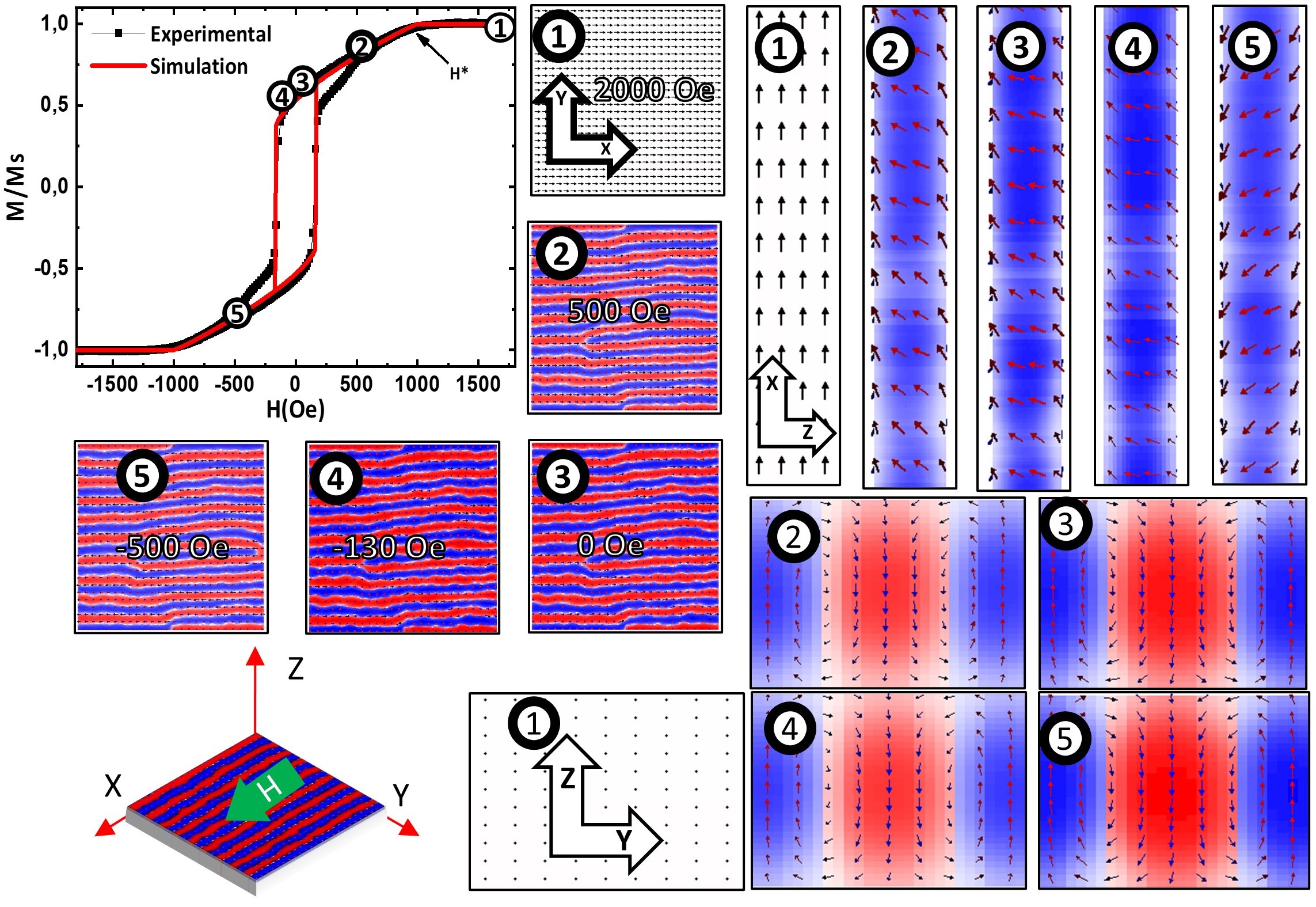}
\caption{\label{Sim60} Experimental and simulated hysteresis loops of a FePt 60 nm/Si.  (1-5) Domain structure as a function of magnetic field. Detail of the evolution of the magnetization within a stripe and y-z crossection as a function of magnetic field.}
\end{figure}

Micromagnetic simulations of magnetization loops and the domain structure were performed to gain additional information. The magnetization loops of 60 nm-thick films were reproduced setting a perpendicular magnetic anisotropy constant of $K_{PMA}=1.6\ \mathrm{\frac{Merg}{cm^3}}$ and considering an intergrain interaction of 20\% of $A_{ex}$, the interaction between neighboring spins (Figure \ref{Sim60}). A good agreement between the experimental and the simulated loops is shown in Figure \ref{Sim60}. We attribute the difference between experimental and simulated loops in the region $H_c<H<H^*$ to the presence of defects and edge effects.

The value of the anisotropy constant agrees with the one calculated from MFM measurements in Section \ref{tc_section}, and the reduction of the exchange interaction at the grains boundaries is expected for polycrystalline films \cite{ramos2009stripe, hughes1983magnetization}. For $H<1200$ Oe, the cross sections of the simulated magnetic domains in the x-y plane successfully reproduce the stripes observed in MFM images (Figure \ref{Sim60}). The field dependence of the x-z cross-sections puts in evidence that the magnetization within the stripes reverses by uniform rotation. The magnetic moments at the y-z cross-section arrange to minimize the magnetostatic energy. Closure domains are at surface zones, while vortex cores along the x-axis are observed at the center of the sample ($z\sim t/2$).

\begin{figure}[ht]
\centering
\subfloat{\includegraphics[width=0.5\textwidth]{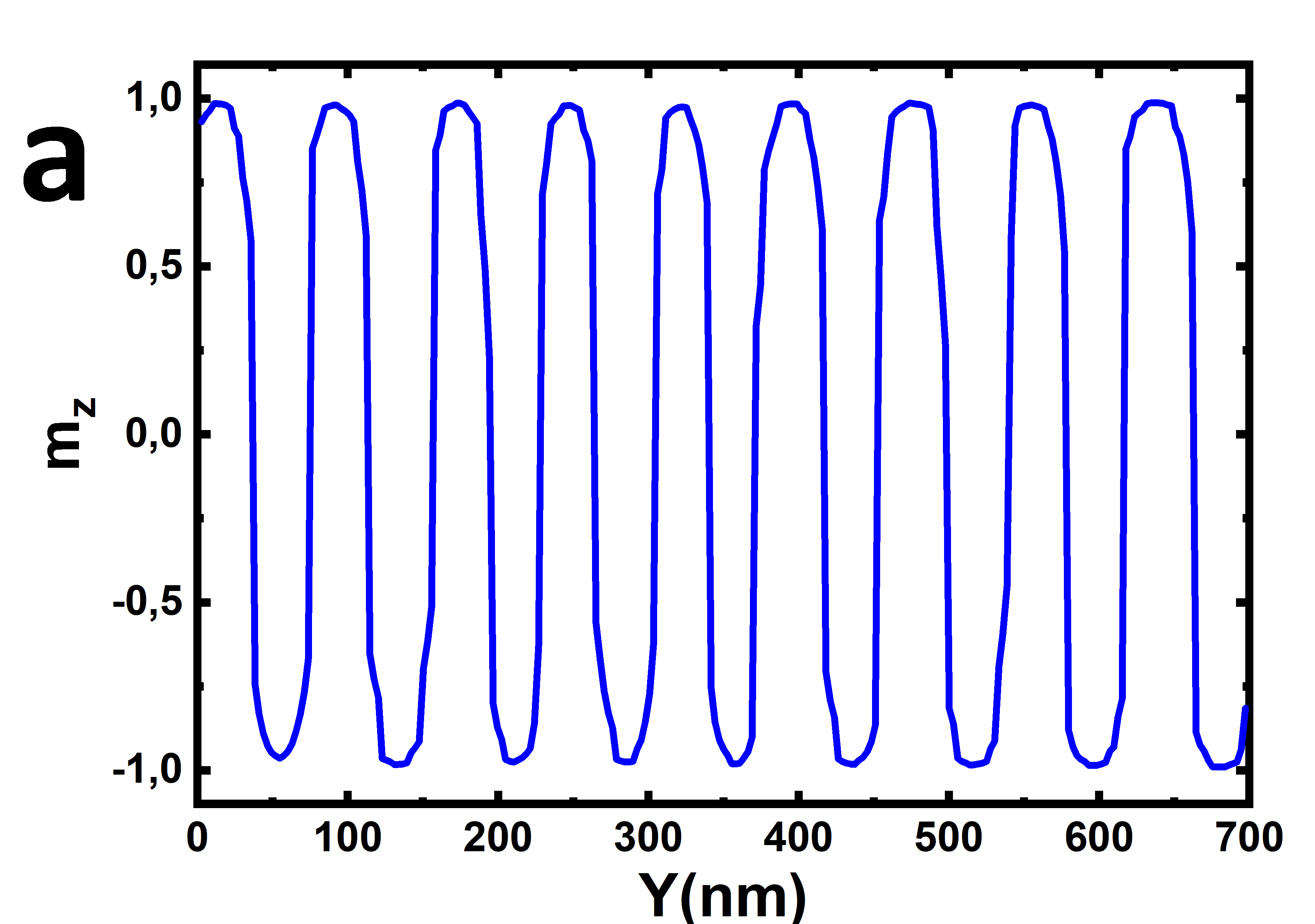}}
\subfloat{\includegraphics[width=0.5\textwidth]{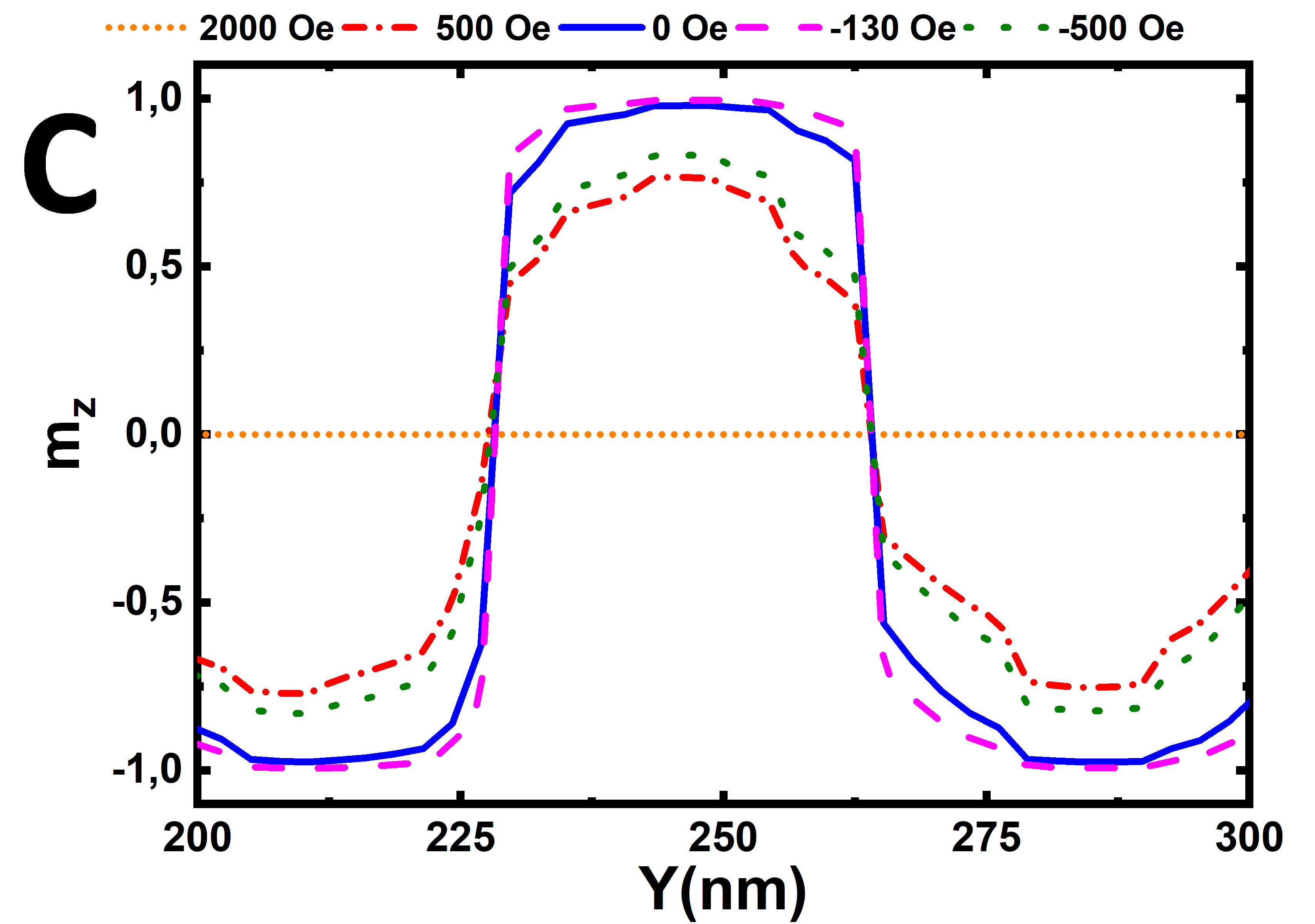}}\\
\subfloat{\includegraphics[width=0.5\textwidth]{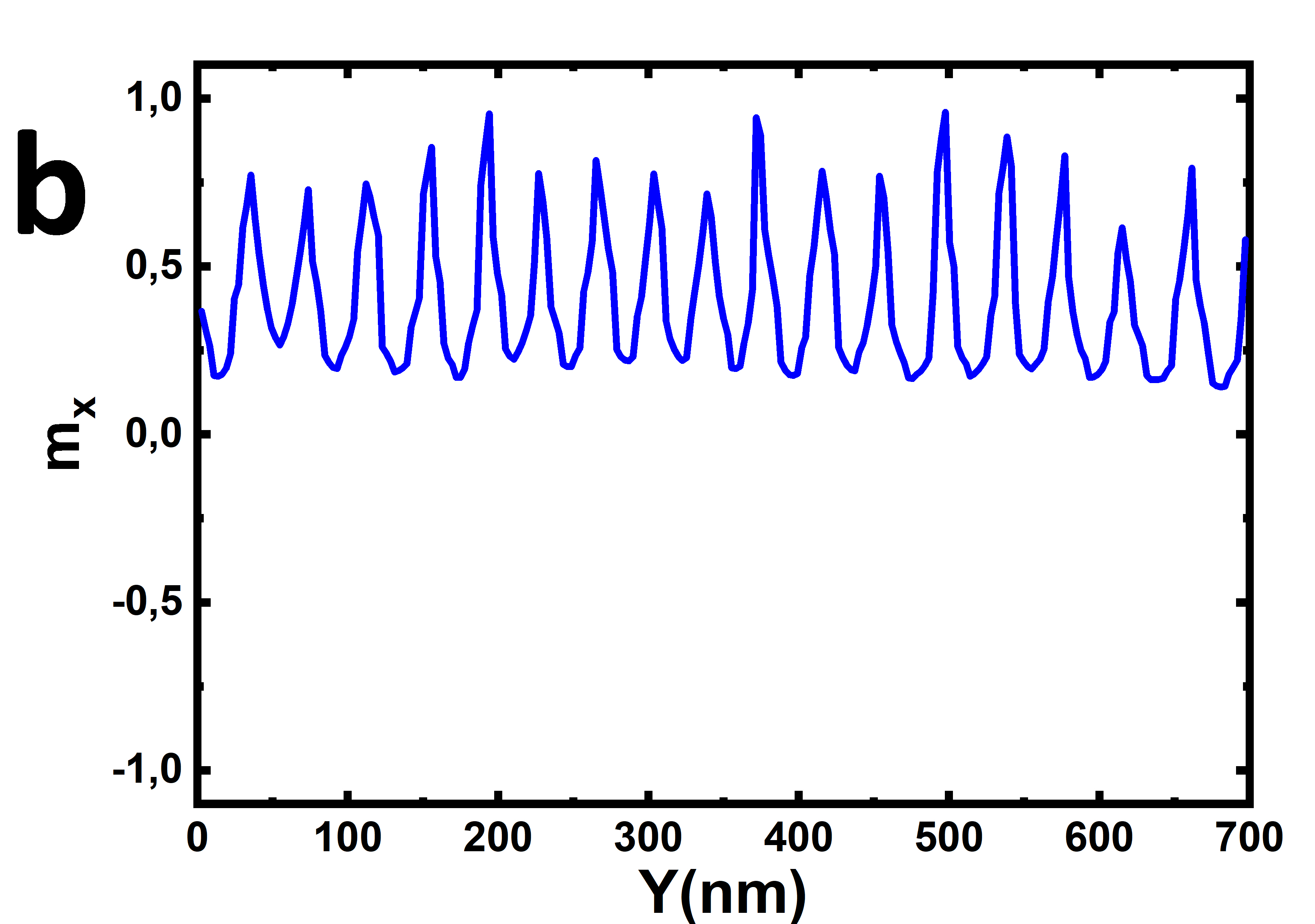}}
\subfloat{\includegraphics[width=0.5\textwidth]{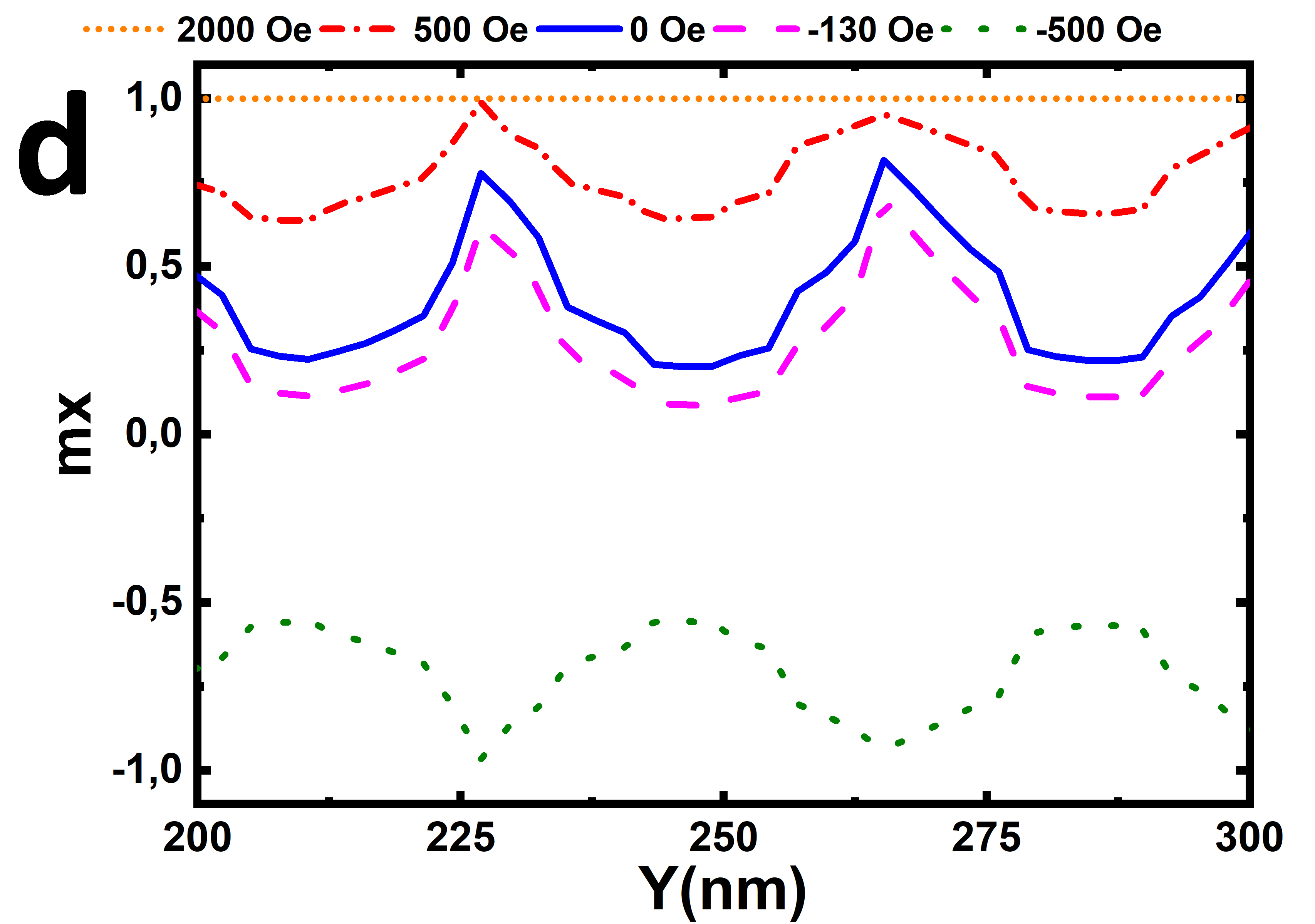}}
\caption{\label{perfil_mz_mx} (a) $m_z$ and (b) $m_x$ profile along Y axis for a 60 nm FePt thin film simulated at remanence. Detail of the magnetic field dependence of the (c) $m_z$ and (d) $m_x$ profile.}
\end{figure}

The profile of the $m_z$ (Figure \ref{perfil_mz_mx} (a) and (c)) component for the 60nm-thickness simulation varies as a sine function along the y-axis, with a period $\lambda_s=90\pm10$ nm that does not depend on the magnetic field. However, $m_z$ reaches the maximum at the coercive field e.g. the magnetic moments rotate out of the plane of the film during the magnetization reversal process. The profile of the $m_x$ component (Figure \ref{perfil_mz_mx} (b) and (d)) presents a periodic dependence with a period of $\frac{\lambda_s}{2}$. For magnetic fields smaller than $H^*$, the amplitude of its oscillations increases as the field decreases until the coercive field ($H_c$), where $m_x$ reverses. 

\subsection{Magnetization reversal at low temperatures}

\begin{figure}[ht]
\centering
\includegraphics[width=\textwidth]{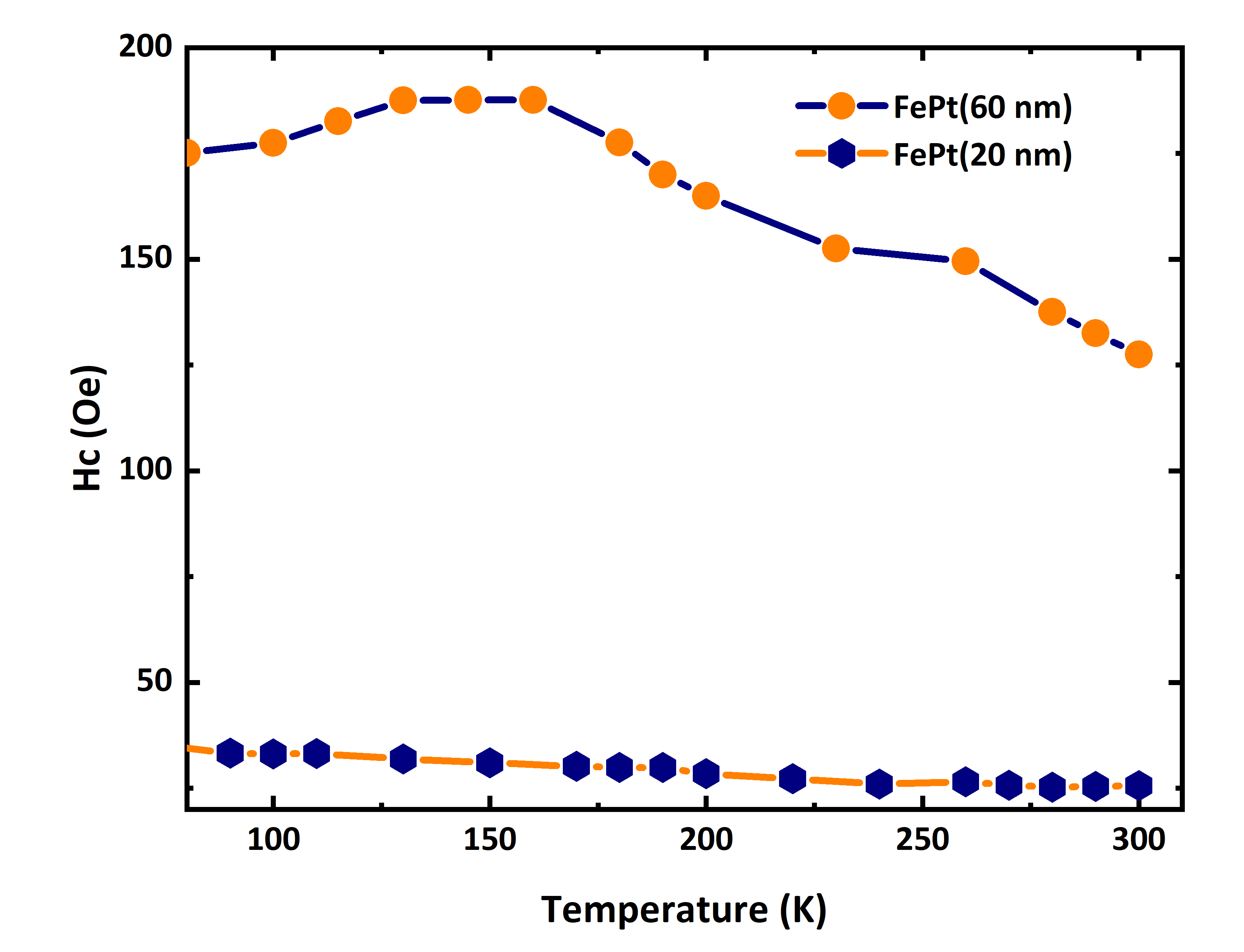}
\caption{\label{Hc_T}Temperature dependence of the coercive field for 60 and 10 nm thickness FePt films.
}
\end{figure}

The coercivity of the thin films ($t<t_c$) decreases with temperature, as a typical ferromagnet does (Figure \ref{Hc_T} ), while the coercive field of thick films ($t>t_c$), instead, presents an abnormal temperature dependence with a maximum located at a characteristic temperature, T*, which has been reported to be thickness dependent by Guzman and coauthors.\cite{guzman2013abnormal}  In their paper, these authors suggested that at T*, there could be a transition in the domain configuration, from a striped-type to fully in-plane domains. 

\begin{figure}[ht]
\centering
\includegraphics[width=\textwidth]{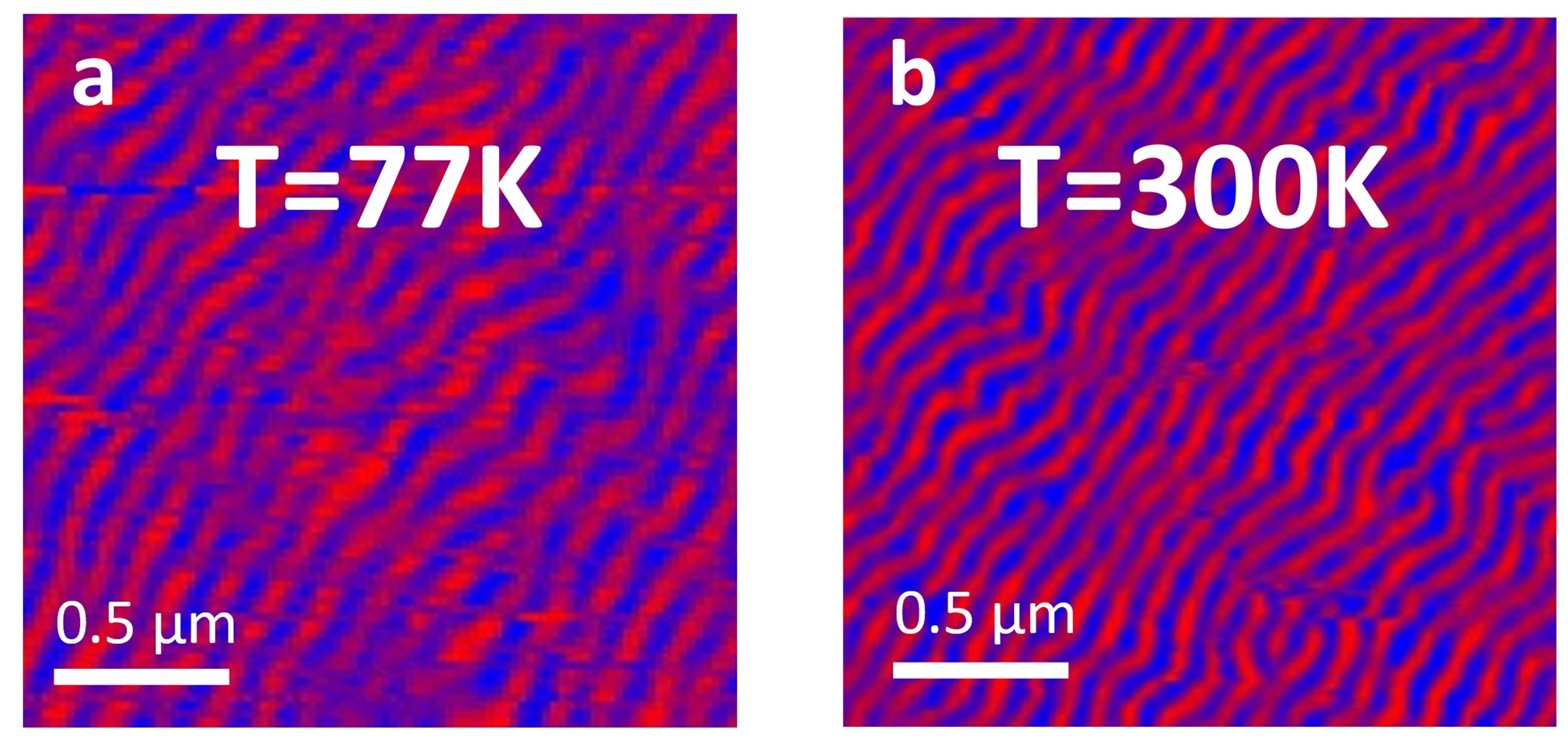}
\caption{\label{MFM_T}MFM images measured at (a) T=77 K and (b)T=300 K for a representative FePt film.
}
\end{figure}

In Figure \ref{MFM_T}, we show images measured by MFM at different temperatures for a 49 nm thick FePt film, which has a maximum coercitivity at $T = 150$ K \cite{guzman2013abnormal}. The domain configuration remains striped between 77 K and room temperature despite the variation of the coercive field behavior as a function of temperature. The stripe period of our samples increases an 18\% as the temperature decreases from 300 to 77 K.  According to Equation \ref{MurayamaLambda}, an increase of the stripes period is associated to a decrease of  the quality ratio Q. Therefore, the increase of $\lambda$ observed in our experiments might indicate a decrease of $K_{PMA}$ as the temperature is lowered. 

\begin{figure}[ht]
\centering
\includegraphics[width=\textwidth]{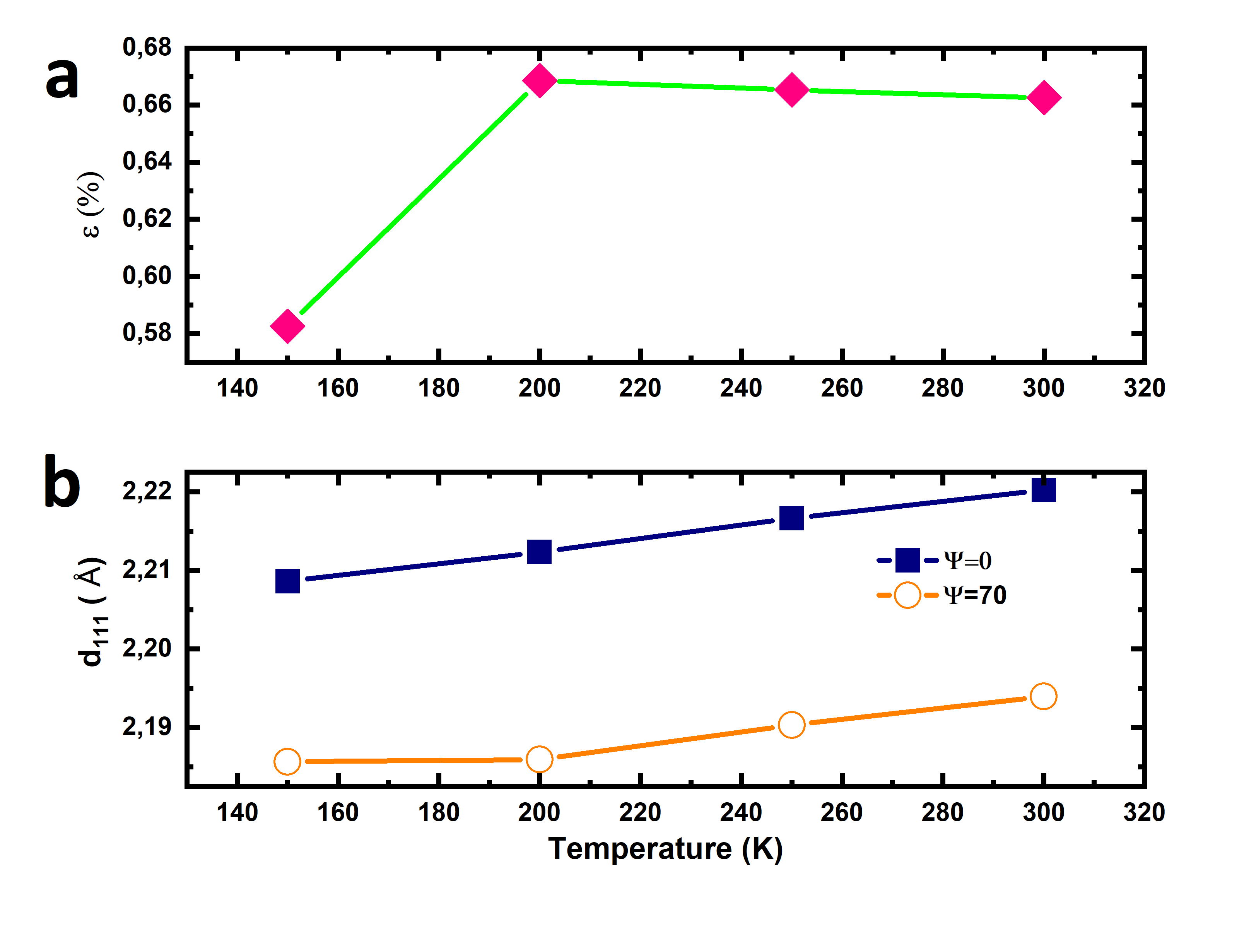}
\caption{\label{d_111_vs_T} (a) d(111) temperature dependence for different XRD geometries. (b) difference between nearly inplane interplannar distance and out of plane interplannar distance.
}
\end{figure}

The change of the perpendicular anisotropy with temperature was associated to magnetostrictive effects. The difference between film and substrate thermal expansion coefficients may induce strains on the films that would change the perpendicular anisotropy \cite{coey2010magnetism, janssen2007stress}. To evaluate the effect of temperature on the strain of the film, we performed XRD measurements below room temperature. We estimated the in-plane compression of d(111) interplane distances from XRD patterns, measured between 150 K and 300 K. The compression $(\epsilon)$ as a function of temperature shown in  Figure \ref{d_111_vs_T} indicates that  there is a reduction of the in-plane strain when the temperature decreases. 
 An anisotropy induced by strains was estimated after the XRD results and using known constants for the FePt system. For the calculation of the stress a Poisson's modulus $\nu=0.33$ and a Young's modulus $E=180$ GPa of FePt \cite{rasmussen2005texture} were used. The magnetostriction coefficient reported for FePt films in the disordered phase is between 100 ppm and 170 ppm \cite{leiva2022electric,ruffonilocal}. The variation of the anisotropy was so calculated, finding a change of $\Delta K_{300\rightarrow{}150}=-0.5±0.1$ $\mathrm{\frac{Merg}{cm^3}}$ when decreasing temperature from RT to 150 K. 
 
\begin{figure}[ht]
\subfloat{\includegraphics[width=0.8\textwidth]{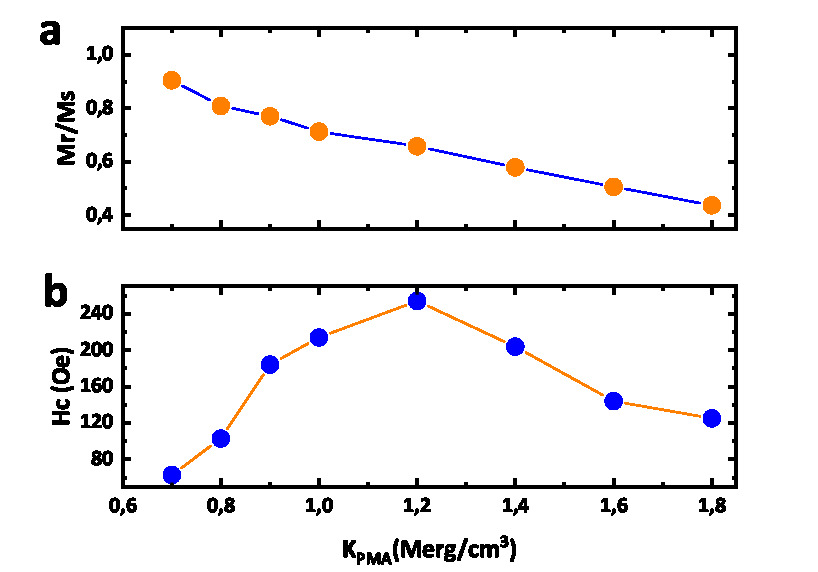}\hspace{-1cm}}
\begin{minipage}[b]{0.25\linewidth}

\includegraphics[width=\textwidth]{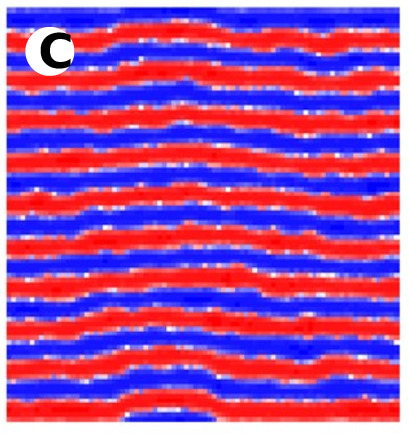}\vspace{-0.1cm}\\
\includegraphics[width=\textwidth]{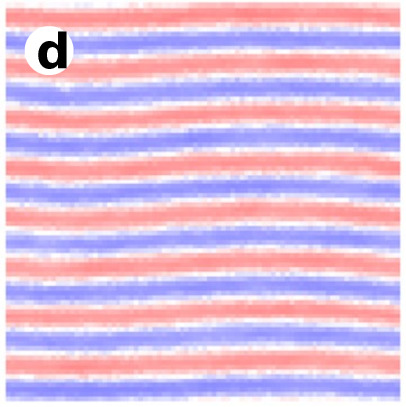}
\vspace{0.2cm}
\end{minipage}
\caption{\label{Sim_vs_KPMA}Simulated (a) remanence and (b) coercive field as a function of the magnetic anisotropy constant for a 60 nm FePt film. Simulated magnetic domain configuration for (c) $K_{PMA}=0.9$ $\mathrm{\frac{Merg}{cm^3}}$ and (d) $K_{PMA}=1.6$ $\mathrm{\frac{Merg}{cm^3}}$.}
\end{figure}

 The influence of a $K_{PMA}$ variation, between 0.7 $\mathrm{\frac{Merg}{cm^3}}$ and 1.8 $\mathrm{\frac{Merg}{cm^3}}$, on the magnetization process was then analyzed by simulations.   Figure \ref{Sim_vs_KPMA} (a) and (b) shows the remanence and coercive field dependence on $K_{PMA}$. The remanence decreases monotonically with increasing $K_{PMA}$ while the coercive field presents a maximum at $K_{PMA}=1.2$ $\mathrm{\frac{Merg}{cm^3}}$. As we show in Figure \ref{Sim_vs_KPMA} (c) and (d), simulated domains are organized in stripes in the whole $K_{PMA}$ range.  The stripes period increases 20\% as the anisotropy constant decreases a 60\%. The change of the coercivity with anisotropy, shown in Figure \ref{Sim_vs_KPMA}, proves that the anisotropy variation with temperature is enough for the system to change its coercivity  through its maximum value.

\begin{figure}[ht]
\centering
\includegraphics[width=\textwidth]{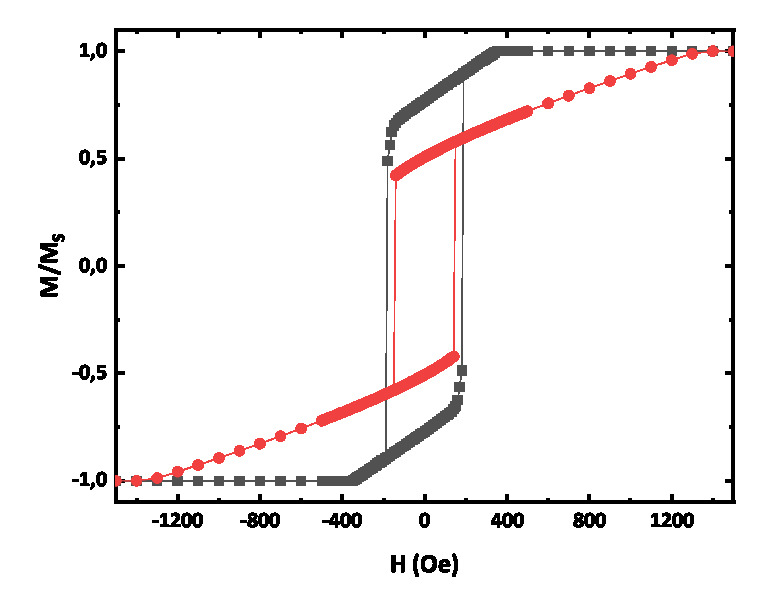}
\caption{\label{ciclos_sim_KPMA}Simulated hysteresis loops from a 60 nm FePt film with ($\blacksquare$) 0.9 $\mathrm{\frac{Merg}{cm^3}}$ and (\textcolor{red}{$\bullet$}) 1.6 $\mathrm{\frac{Merg}{cm^3}}$ perpendicular magnetic anisotropy constant.
}
\end{figure}

 In Figure \ref{ciclos_sim_KPMA}, we present hysteresis loops calculated for anisotropy values higher and lower than $K_{PMA-MAX}=1.2$ $\mathrm{\frac{Merg}{cm^3}}$, where the maximum coercivity is observed. The loop calculated for $K_{hi}=1.6 \mathrm{\frac{Merg}{cm^3}}>K_{PMA-MAX}$ presents a sharp change in the magnetization at the coercive field. In contrast, the one calculated for $K_{lo}= 0.9\mathrm{\frac{Merg}{cm^3}}<K_{PMA-MAX}$ exhibits a smooth variation of the magnetization near the coercive field.
 
\begin{figure}[ht]
\centering
\includegraphics[width=\textwidth]{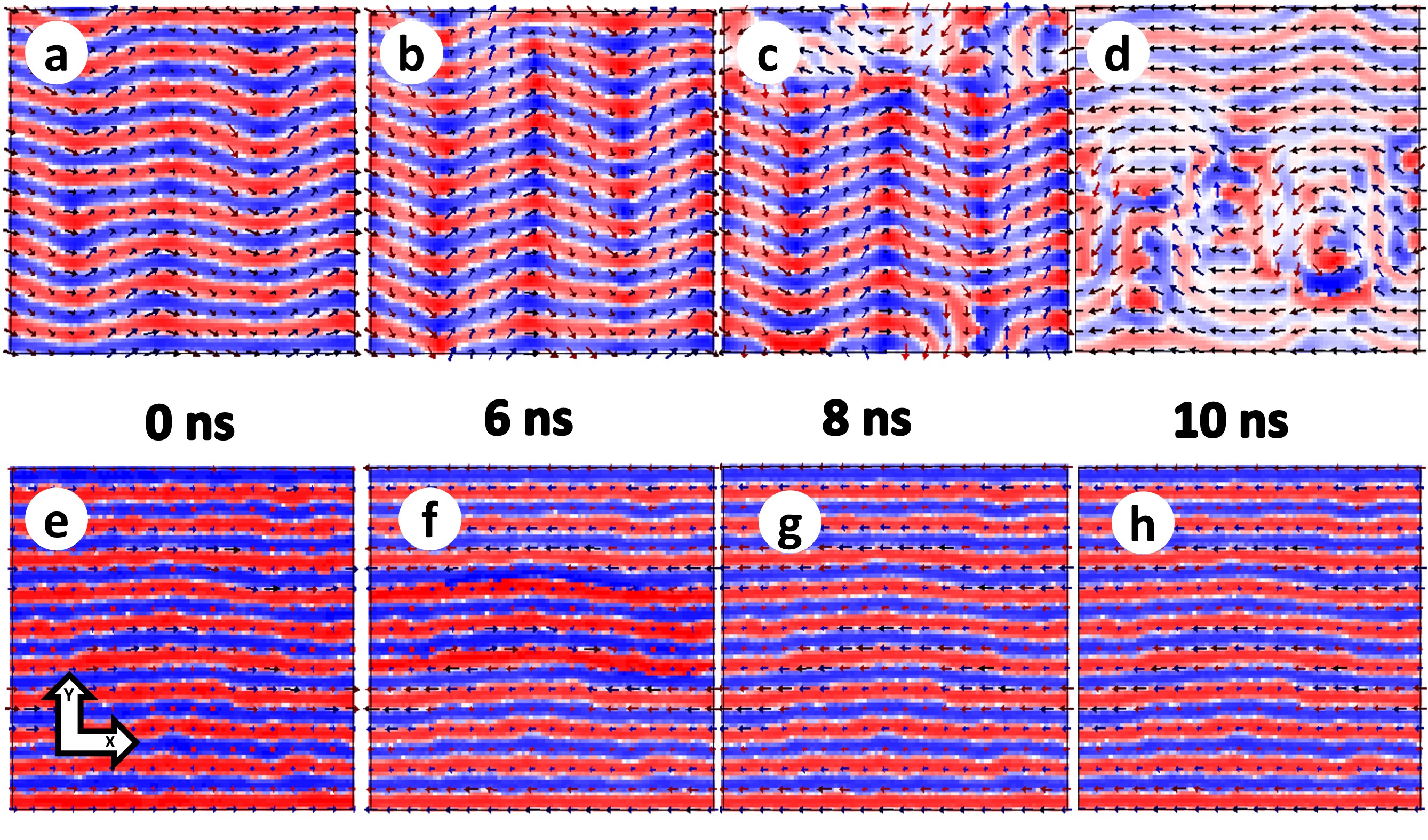}
\caption{\label{Inversion_SIM_vs_t}Simulated time evolution of the magnetic domain configuration from a 60 nm FePt film with (a-d) $K_{PMA}=0.9$ $\mathrm{\frac{Merg}{cm^3}}$ and (e-h) $K_{PMA}=1.6$ $\mathrm{\frac{Merg}{cm^3}}$.}
\end{figure}

 In order to understand the difference in the reversal magnetization process of the calculated loops, we simulated the dynamic of the micromagnetic state near the coercive field for both cases. The initial magnetization state (t=0ns) was taken to be that of M($H^{*} \lesssim H_{c}$) calculated for $K_{lo}$ and $K_{hi}$(Figure \ref{Inversion_SIM_vs_t}  (a) and (e)).  At t$>$0 ns, a magnetic field $H^{**}$  slightly larger than the coercivity was applied to the systems and the temporal evolution of the magnetic state was simulated for $K_{lo}$ (Figure \ref{Inversion_SIM_vs_t}  top) and $K_{hi}$ (Figure  \ref{Inversion_SIM_vs_t}  bottom). Both $H^{*}$ and $H^{**}$ were applied along the -$x$ direction.

For $K_{lo}$, the magnetic moments rotate in the plane of the film. During the first 6 ns, the spins rotate from the x-axis to the y-axis (Figure \ref{Inversion_SIM_vs_t} (a) and (b)). At t= 8ns, the formation of vortices gave place to the appearance of reversed domains parallel to the applied field (Figure \ref{Inversion_SIM_vs_t} (c)). Finally, after 10 ns, the formed domains expand from one side of the sample to its center (Figure \ref{Inversion_SIM_vs_t} (d)). On the other side, For $K_{PMA}>K_{PMA-MAX}$, the magnetic moments rotate out of the plane of the film. The stripe structure remains throughout the whole reversal magnetization process. There is a coherent rotation of the magnetic moments of each stripe, but the stripes don't rotate simultaneously.

The reversal magnetization process is notably faster for films with $K_{hi}$. As shown in Figure \ref{Inversion_SIM_vs_t}, the spins complete their rotation after 8 ns. In comparison, the reversal process ends long after 10 ns for films with $K_{lo}$.   The simulation results proved that the abnormal temperature dependence of the coercive field is related to a variation of the perpendicular magnetic anisotropy induced by the mismatch between thermal expansion coefficients of films and substrate,  inducing a change of the reversal magnetization process.

The existence of a magnetic anisotropy constant value where the coercivity reaches a maximum ($K_{PMA-MAX}$) was also observed in simulations of FePt films with thicknesses between 30 nm and 60 nm. Guzmán and co-workers \cite{guzman2013abnormal} report that when the thickness of the film decreases, the maximum in coercitivity occurs at temperatures closer to room temperature. In agreement with this result, we observed that as the thickness of the simulated film decreases, the value of $K_{PMA-MAX}$ increases. Also, a maximum in the thickness dependence of the coercivity was observed, in agreement with the results published by Sallica and co-workers\cite{leva2010magnetic}.

\section{Conclusions}

We have performed a detailed study of the magnetization reversal mechanisms in FePt films with moderate PMA using micromagnetic simulations and experiments. These films show a transition from planar to stripe-like magnetic domains above a critical thickness $t_c$. 
At room temperature, the reversal mechanism of the thinner films ($t<t_c$) was described by the two-phase model that combines coherent rotation with domain-wall movement. In contrast, the striped domain configuration,  observed at remanence,  dominates the magnetization reversal process in thicker films ($t>t_c$). Moreover,  we succeded in explaining the origin of the anomalous temperature variation of the coercive field observed  in thick  films. The effect was associated to a change of the magnetization reversal process dynamic due to the variation of a PMA by substrate-induced strains detected in the structural characterization of the samples.  

\section*{Acknowledgments}
Authors thanks the financial support of FONCYT PICT 0867-2016, PICT 02781-2019 and the European Commission through the Horizon H2020 funding by H2020-MSCA-RISE-2016 - Project N$^0$ 734187 –SPICOLOST. Beamtime was granted on the XRD2 beamline by the LNLS.
C.G. acknowledges the financial support received by ANID FONDECYT/REGULAR 1201102, ANID PIA/APOYO AFB180002 and ANID FONDEQUIP EQM140161.
\bibliography{References}
\newpage
\clearpage
\onecolumngrid
\section*{Figures}

\
\
\
\
\
\
\
\
\
\
\
\
\
\
\clearpage
\end{document}